\documentclass[a4paper,fleqn,usenatbib]{mnras}

\usepackage[T1]{fontenc}
\usepackage{ae,aecompl}

\usepackage{graphicx}	% Including figure files
\usepackage{amsmath}	% Advanced maths commands
\usepackage{amssymb}	% Extra maths symbols

\title[Proper Motion of Sextans Dwarf Galaxy]{Proper Motion of the Sextans Dwarf Galaxy from Subaru Suprime-Cam Data}

\author[D. I. Casetti-Dinescu \& T. M. Girard]{
Dana I. Casetti-Dinescu,$^{1}$\thanks{E-mail: danacasetti@gmail.com}
Terrence M. Girard,$^{2}$
\thanks{Email: terrence.girard@gmail.com}
Michael Schriefer$^{1}$
\\
$^{1}$Physics Department, Southern Connecticut State University, 
New Haven, CT 06515-1355, USA \\
$^{2}$ 14 Dunn Road, Hamden CT 06518, USA
}

\date{Accepted XXX. Received YYY; in original form ZZZ}

\pubyear{2017}

\begin{document}
\label{firstpage}
\pagerange{\pageref{firstpage}--\pageref{lastpage}}
\maketitle

\begin{abstract}
We have measured the absolute proper motion of the Sextans dwarf spheroidal 
galaxy using $Subaru$ Suprime-Cam images taken at three epochs, with a time baseline of $\sim 10$ years. We astrometrically calibrate each epoch by constructing
distortion-correction ``maps'' from the best available $Subaru$ Suprime-Cam
dithered data sets and from $Gaia$ DR1 positions. The magnitude limit of the proper-motion study is $V \sim 24$. The
area covered is $26\farcm7\times23\farcm3$, which is still within the core radius of Sextans. The derived proper motion is 
$(\mu_{\alpha}, \mu_{\delta}) =(-0.409\pm0.050, -0.047\pm0.058)$ mas yr$^{-1}$.  The direction of motion is perpendicular
to the major axis of the galaxy. Our measurement, combined with
radial velocity and distance from the literature, 
implies a low eccentricity orbit, with a moderate inclination to the Galactic plane, and a period of 3 Gyr. Sextans
is now some 0.4 Gyr away from its pericenter ($r_{peri} \sim 75$ kpc), moving toward its apocenter ($r_{apo} \sim 132$ kpc).
Its orbit is inconsistent with membership to the vast polar structure of Galactic satellites. 
\end{abstract}



%%%%%%%%%%%%%%%%%%%%%%%%%%%%%%%%%%%%%%%%%%%%%%%%%%

%%%%%%%%%%%%%%%%% BODY OF PAPER %%%%%%%%%%%%%%%%%%

\section{Introduction}
Dwarf-galaxy satellites of the Milky Way (MW) --- 
composed of the eleven classical dwarf galaxies along with the 
thirty plus galaxies more recently discovered and including the ultra-faint dwarfs ---  display a wide range of properties. 
Nevertheless, two major traits
appear in all of these systems: 1) the majority of their stars are basically old and metal poor, and 2)
their internal kinematics is not accounted for by the luminous mass, under equilibrium conditions, implying the existence of 
dark matter. The amount of dark matter, as inferred from the mass-to-light ratios, varies wildely
from very large values in all of the ultra-faint dwarfs \citep[e.g.,][]{si11,sg07} \citep[but see the note of caution by][]{spencer17},
to large values in the classical dwarfs Draco, Ursa Minor and Sextans \citep{mateo98,mcc12}, to 
moderate values in most classical dwarfs. The range in size also varies tremendously, from the 
small ultra-faint dwarfs with half-radii of a couple to a few tens of pc, 
to very extended systems such as Sextans and Fornax with $r_h \sim 700$ pc \citep{mcc12}, to finally,
the recently discovered Crater 2 \citep{torr16}, with a half light radius of 1 kpc. This last system, although displaying unusually
cold kinematics for its luminosity and spatial extent \citep{cald17}, is still dark-matter dominated.

Another property, which is largely not understood, is the morphological appearance in relation to the distance from the MW center. 
Simplistically, it is expected
that satellites that are closer to the MW will exhibit morphology related to tidal interaction, while those that are more distant will not.
Contrary to this expectation, Draco at ~80 kpc, has a smooth, regular morphology \citep{seg07} while Ursa Minor at a similar
distance shows tidal-distortion signatures \citep{pal03,bel02}.
Similarly, Leo IV at 160 kpc shows no signs of tidal interaction \citep{san10}, 
while Leo V at 180 kpc shows overdensities at large radii from its center \citep{san12}.
Finally, Hercules at $\sim 140$ kpc shows a very elongated morphology, with overdensities some 1.9 kpc from the galaxy \citep{rode15}, 
which has led \citet{kup17} to suggest that the satellite is on a nearly-radial orbit.

Clearly, more work needs to be done if a consistent and coherent picture of the MW's satellite system and its formation is to be
accomplished. Here, we focus on Sextans, a classical dwarf spheroidal galaxy discovered in 1990 \citep{irw90} from the UK Schmidt
photographic survey. Sextans is old and metal poor, with a rather high mass-to-light ratio \citep{mcc12}; tidal stirring has been
suggested as a cause for its large velocity dispersion. Little was known about the morphological structure of Sextans, given 
its extended size (half-light radius $\sim 26\arcmin$), until \citet{rode16} constructed a wide map of the galaxy using
the Dark Energy Camera on the Blanco 4m telescope. They found overdensities within the tidal radius of the galaxy, and suggest
that Sextans is not under strong tidal stress from the MW. In the sky, Sextans is located within the vast polar structure (VPOS)
outlined by \citep[][see also references therein]{paw13,paw15}, in which a large number of classical dwarfs and ultra faint dwarfs lie.  
Arguably, this structure poses challenges to cosmological model predictions \citep{paw15}.  It is, thus, crucial to measure
the proper motions of these systems, in order to determine whether this structure is also kinematically disk-like. 

For these reasons, we have undertaken
to measure the absolute proper motion of Sextans, a system that has no such astrometric measurement to date. The only
estimate of its proper motion comes from radial velocities under the assumption that the gradient in these velocities 
is due to the projected transverse motion. This indirect estimate, made by \citet{wal08}, indicates a motion away from the VPOS;
although the large uncertainty, due to the low velocity gradient, renders the determination inconclusive.

Here we use archived images taken with a wide-field camera, the Suprime-Cam on the {\it Subaru} 8m telescope, to measure the proper motion
of Sextans. We develop a methodology to calibrate the camera, not only to exploit the astrometric potential of this instrument, but also
as a testbed for the Dark Energy Camera surveys as well as the upcoming Large Synoptic Survey Telescope (LSST). 
The accuracy of this type of study is typically limited by the the proper-motion zero-point determination,
which derives from the large number of background galaxies available for measurement
and, in the near future, from foreground
stars that will overlap with the magnitude coverage of the upcoming {\it Gaia} mission data releases. In short, this type of
work is a step toward extending the depth of wide-field proper-motion studies, with direct application to distant and/or
sparse, low surface-brightness systems orbiting the MW. This is the second paper in our series; the first 
focused on the Draco dwarf galaxy \citep{cg16} and used a slightly different astrometric calibration procedure than what is presented here. 
The novelty of this work consists of the calibration of the data at each epoch, either internally from overlapping offsets, or
externally, using {\it Gaia} DR1 \citep{gaia16}.

\section{Observations}

The $Subaru$ Mitaka Okayama Kiso Archive system (SMOKA), \citep{baba02}
was used to search for Suprime-Cam \citep{miy02} images containing the Sextans
dwarf galaxy. We have found data taken in 2002, 2003, 2004, 2005, 2006, 2012 and 2013.
The 2004 data set was rather shallow and of 
poor quality, therefore we have discarded it from our analysis. The 2013 data set included only three exposures taken 
at a single pointing, far ($24\arcmin$) from the center of Sextans and in a non-traditional $VR$ filter; we therefore
 chose not to use it in our analysis. We are thus left with an early epoch consisting of 2002-2003 data, an
intermediate epoch including 2005-2006 data, and a modern epoch at 2012.

In addition to our primary target, Sextans, we have also searched 
the archive for data sets with many offsets ($ \ge 10\arcmin$) and dithers
(a few arcmin) in dense stellar fields, observations using similar filters and taken near the epochs 
of those for Sextans. Such data sets are needed in conjunction with the Sextans data to astrometrically map out the 
Suprime-Cam at each epoch.
A series of data sets centered on globular cluster NGC 2419 was found that matches these
criteria well. This field is also very interesting astrophysically; it includes the distant and massive globular cluster NGC 2419 
whose proper motion was first measured recently based on $Gaia$ DR1 and $HST$
\citep{mass17}, and the Monoceros stellar overdensity in the foreground
of the cluster \citep{carb15}. Our proper-motion results for this field will be
presented in a future paper.

The total data set for Sextans consists of 246 individual exposures; the characteristics of these exposures are listed in Table~\ref{tab:tab1}.
Exposures in the early and modern epochs are aligned with the equatorial system, while those of the intermediate-epoch are 
rotated by $34\degr$ and $37\degr$. 

\begin{figure}
\includegraphics[width=\columnwidth,angle=0]{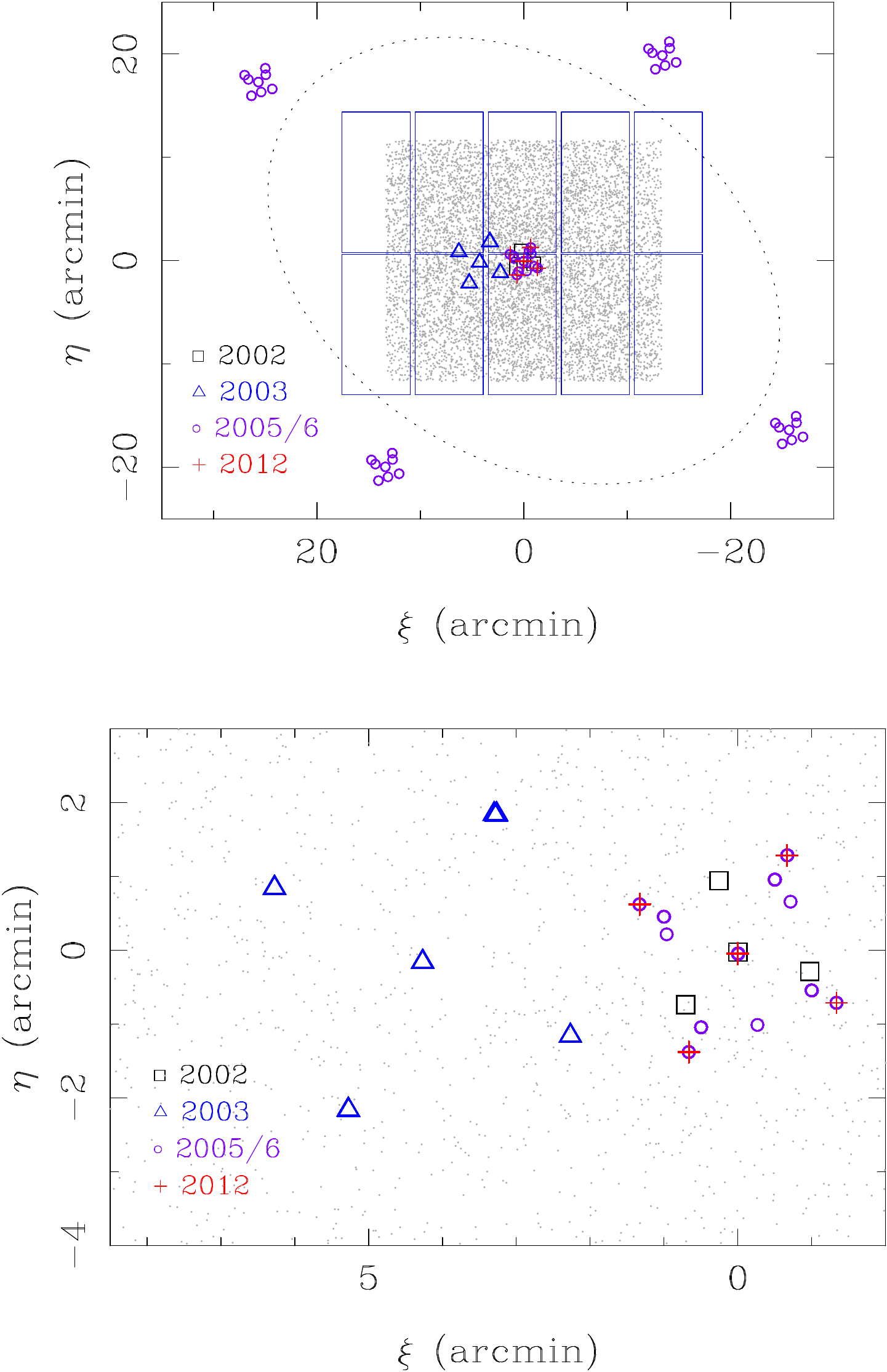}
\caption{
Spatial distribution of the Sextans observations. Field centers of the Subaru exposures
are shown with large symbols, colour- and shape-coded by epoch of observation.
Grey points show objects from our final proper-motion catalogue.
North is up, and East to the left. The origin of the plot is at the 
center of Sextans, and the ellipse shows the extent of the core of the dwarf galaxy.
The wireframe illustrates a single exposure's field of view.
The ``zoomed-in'' bottom panel shows details at the center of the top panel.}
\label{fig:fig1}
\end{figure}

We show the spatial coverage of the Sextans data in Figure~\ref{fig:fig1}, in standard coordinates ($\xi,\eta$).
The origin is at the dwarf galaxy's center as adopted from \citet {irw90}. 
The large symbols show the pointings of the individual Suprime-Cam exposures
while the grey dots show objects from our final proper-motion catalog, the latter being confined to the area
that is best-calibrated by our procedures. 
Not all 2005/6 observations from Table~\ref{tab:tab1} are
shown in Figure~\ref{fig:fig1}, due to the area restriction of the plot. The full areal extent of these observations are
presented in Section 6.1, where construction of the distortion maps is described. 
Suprime-Cam is a mosaic of ten $2048\times4096$ CCDs located at the prime focus of the $Subaru$ telescope.
It covers a $34\arcmin\times27\arcmin$ field of view, with a pixel scale of $0\farcs20$ \citep{miy02}. 
The rectangular wireframe shown in Figure~\ref{fig:fig1} outlines the CCDs of a sample exposure. 
The ellipse indicates the extent of the core of 
Sextans with semimajor axis, ellipticity, and position angle taken from \citet{rode16}.
\begin{table}
\caption{Observation Log}
\label{tab:tab1}
\begin{tabular}{llll}
\hline
\multicolumn{1}{c}{Date} & \multicolumn{1}{c}{Filter} & \multicolumn{1}{c}{Exp. time} & \multicolumn{1}{c}{N$_{exp}$} \\
 & & \multicolumn{1}{c}{(sec)} & \\
\hline
\multicolumn{1}{c}{2002/11/07} & \multicolumn{1}{l}{$V$} & \multicolumn{1}{r}{10} & \multicolumn{1}{r}{1} \\
\multicolumn{1}{c}{''} & \multicolumn{1}{l}{''} & \multicolumn{1}{r}{240} & \multicolumn{1}{r}{3} \\
\multicolumn{1}{c}{''} & \multicolumn{1}{l}{$I_C$} & \multicolumn{1}{r}{10} & \multicolumn{1}{r}{1} \\
\multicolumn{1}{c}{''} & \multicolumn{1}{l}{''} & \multicolumn{1}{r}{80} & \multicolumn{1}{r}{2} \\
\multicolumn{1}{c}{2003/04/01,03} & \multicolumn{1}{l}{$V$} & \multicolumn{1}{r}{10} & \multicolumn{1}{r}{5} \\
\multicolumn{1}{c}{''} & \multicolumn{1}{l}{''} & \multicolumn{1}{r}{240} & \multicolumn{1}{r}{4} \\
\multicolumn{1}{c}{''} & \multicolumn{1}{l}{$I_C$} & \multicolumn{1}{r}{30} & \multicolumn{1}{r}{5} \\
\multicolumn{1}{c}{''} & \multicolumn{1}{l}{''} & \multicolumn{1}{r}{240} & \multicolumn{1}{r}{5} \\
\multicolumn{1}{c}{\bf Early-epoch total} & \multicolumn{2}{c}{.............} & \multicolumn{1}{r}{{\bf 26}} \\ \\
\multicolumn{1}{c}{2005/02/05} & \multicolumn{1}{l}{$V$} & \multicolumn{1}{r}{230} & \multicolumn{1}{r}{2} \\
\multicolumn{1}{c}{''} & \multicolumn{1}{l}{$I_C$} & \multicolumn{1}{r}{270} & \multicolumn{1}{r}{7} \\
\multicolumn{1}{c}{2005/05/04-05} & \multicolumn{1}{l}{$B$} & \multicolumn{1}{r}{5} & \multicolumn{1}{r}{2} \\
\multicolumn{1}{c}{''} & \multicolumn{1}{l}{''} & \multicolumn{1}{r}{10} & \multicolumn{1}{r}{2} \\
\multicolumn{1}{c}{''} & \multicolumn{1}{l}{''} & \multicolumn{1}{r}{60} & \multicolumn{1}{r}{5} \\
\multicolumn{1}{c}{''} & \multicolumn{1}{l}{''} & \multicolumn{1}{r}{500} & \multicolumn{1}{r}{5} \\
\multicolumn{1}{c}{''} & \multicolumn{1}{l}{$I_C$} & \multicolumn{1}{r}{5} & \multicolumn{1}{r}{2} \\
\multicolumn{1}{c}{''} & \multicolumn{1}{l}{''} & \multicolumn{1}{r}{10} & \multicolumn{1}{r}{2} \\
\multicolumn{1}{c}{''} & \multicolumn{1}{l}{''} & \multicolumn{1}{r}{60} & \multicolumn{1}{r}{5} \\
\multicolumn{1}{c}{''} & \multicolumn{1}{l}{''} & \multicolumn{1}{r}{200} & \multicolumn{1}{r}{15} \\
\multicolumn{1}{c}{2005/12/31-2006/01/01-03} & \multicolumn{1}{l}{$V$} & \multicolumn{1}{r}{50} & \multicolumn{1}{r}{60} \\
\multicolumn{1}{c}{''} & \multicolumn{1}{l}{$I_C$} & \multicolumn{1}{r}{30} & \multicolumn{1}{r}{33} \\
\multicolumn{1}{c}{''} & \multicolumn{1}{l}{''} & \multicolumn{1}{r}{235} & \multicolumn{1}{r}{60} \\
\multicolumn{1}{c}{\bf Intermediate-epoch total} & \multicolumn{2}{c}{.............} & \multicolumn{1}{r}{{\bf 200}} \\ \\
\multicolumn{1}{c}{2012/12/09,16} & \multicolumn{1}{l}{$b$} & \multicolumn{1}{r}{36} & \multicolumn{1}{r}{4} \\
\multicolumn{1}{c}{''} & \multicolumn{1}{l}{''} & \multicolumn{1}{r}{360} & \multicolumn{1}{r}{6} \\
\multicolumn{1}{c}{''} & \multicolumn{1}{l}{$y$} & \multicolumn{1}{r}{18} & \multicolumn{1}{r}{4} \\
\multicolumn{1}{c}{''} & \multicolumn{1}{l}{''} & \multicolumn{1}{r}{180} & \multicolumn{1}{r}{6} \\
\multicolumn{1}{c}{\bf Modern-epoch total} & \multicolumn{2}{c}{.............} & \multicolumn{1}{r}{{\bf 20}} \\
\hline
\end{tabular}
\end{table}

\begin{figure}
\includegraphics[width=\columnwidth,angle=0]{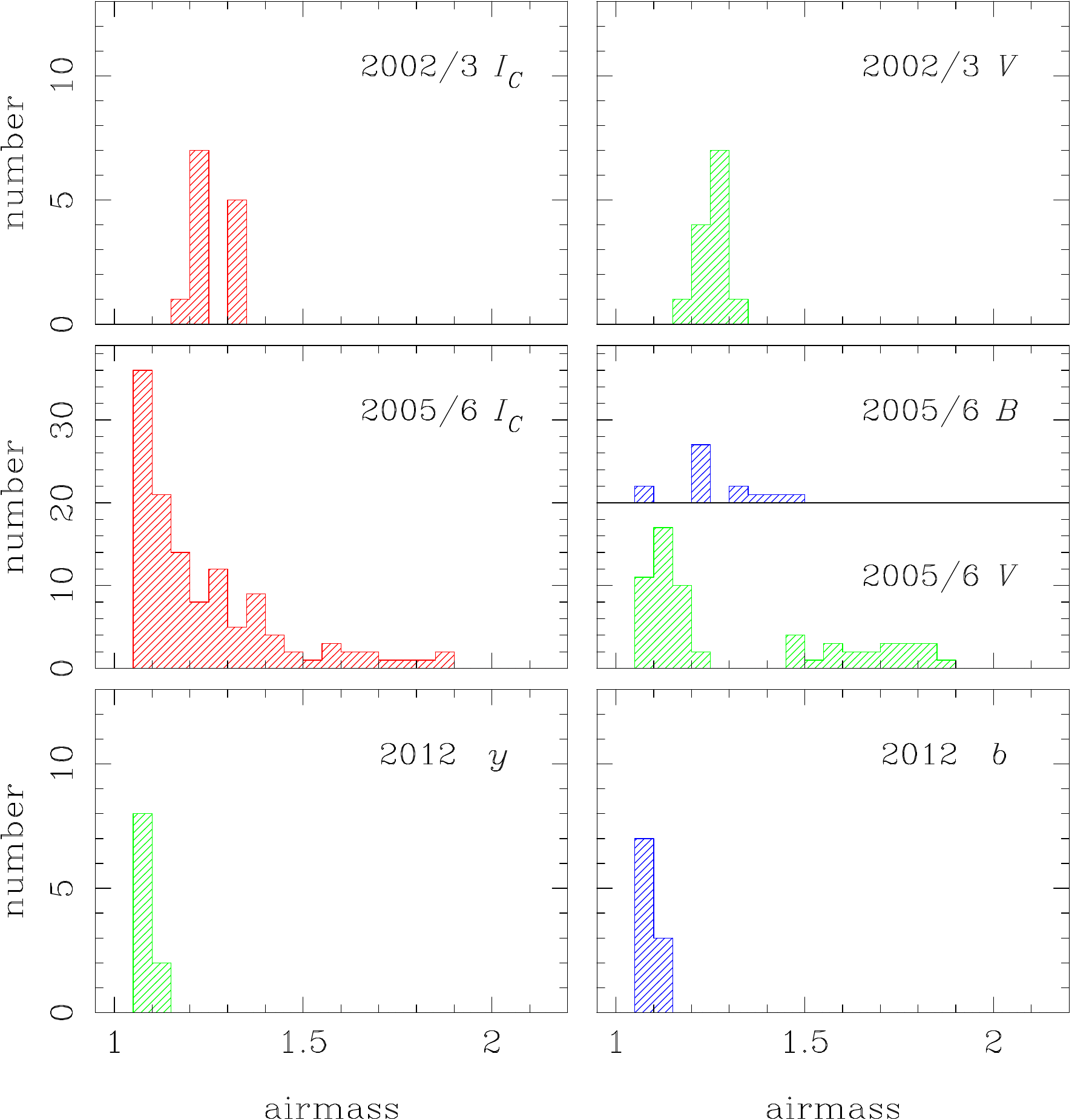}
\caption{Distribution in airmass, per filter and epoch.}
\label{fig:fig2}
\end{figure}

\begin{figure}
\includegraphics[width=\columnwidth,angle=0]{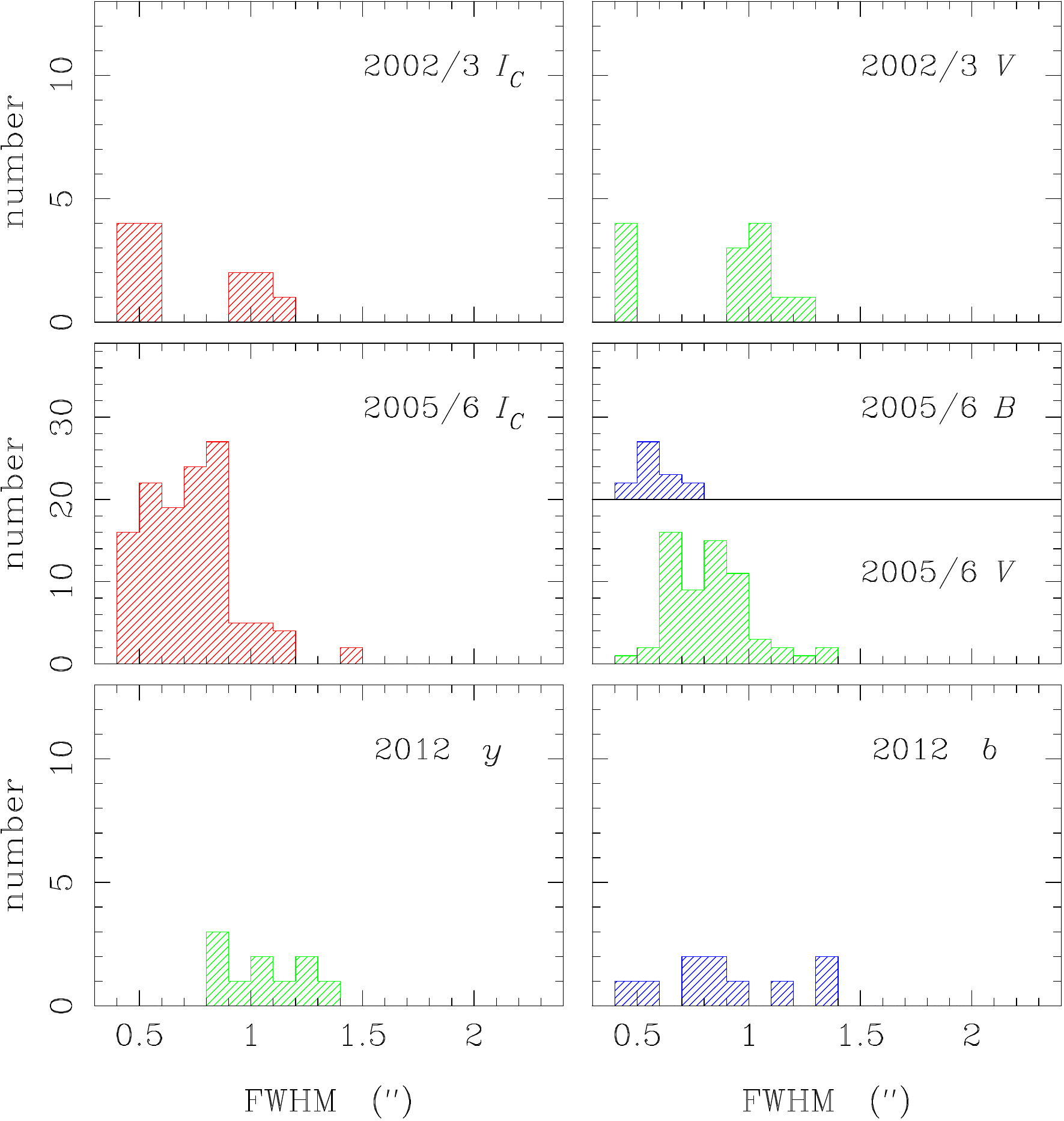}
\caption{Distribution in FWHM, per filter and epoch as determined from one of the central CCDs 
(``si005s'' for the early and intermediate-epoch 
observations and ``satsuki'' for the modern-epoch observations).}
\label{fig:fig3}
\end{figure}

$Subaru$'s prime focus corrector includes an atmospheric dispersion corrector (ADC) used to
correct for chromatic atmospheric dispersion \citep{miy02}. All of the observations considered here were made with the ADC in use.
We show histograms of the airmass distribution for the three epochs and for each filter
in Figure~\ref{fig:fig2}. The 2012 observations were taken closest to the meridian, while those of the intermediate-epoch span
a large range in airmass. In Figure~\ref{fig:fig3} we show the FWHM distributions as determined for one of the central CCDs.
The intermediate-epoch $I_C$-band observations have the best average FWHM, and it is these observations that we will use for object
classification. 

\section{Basic Data Processing}

We treat each CCD of the mosaic Suprime-Cam as a single unit throughout the astrometric and photometric calibration steps.
Pre-processing, including overscan, bias subtraction and flat fielding are performed using the 
package SFRED1 for data before July 21, 2008, (the date the CCD detectors were replaced), and package SFRED2 after this date,
\citep{ya02,ou04}\footnote{www.subarutelescope.org/Observing/Instruments/SCam/sdfred/}. 
We have not included the distortion and differential geometric atmospheric dispersion corrections
available in these standard reduction packages since they apply the corrections directly to the pixel data. 
Instead, we account for these effects during the astrometric solutions, as various plate terms (see
Section 6). The PSF-size estimate from SDFRED1/2 provides a measure of FWHM for 
each exposure and individual CCD, and this we use as input in the detection step.
The code SourceExtractor  \citep{be96} is employed to provide 
object detection, photometry, preliminary centers and object classification. 
Final object centers are determined by fitting a two-dimensional elliptical Gaussian function
to each object's intensity data. We use the Yale centering routines upgraded for CCD data from the original code described in \citet{leeva83}, with the centers from SExtractor serving as initial values for the nonlinear fitting algorithm.

\section{Photometry}
Given the large area covered by the 2005/2006 data, we chose to use SDSS DR9 \citep{ahn12} to calibrate the 
Suprime-Cam instrumental magnitudes. 
These instrumental magnitudes are SExtractor isophotal magnitudes. We use only $V$ and $I_C$-band data from observations
taken in 2002/3 and 2005/6. SDSS magnitudes are transformed to the Johnson system via 
the transformations determined by \citet{jor06}. Each exposure and CCD is calibrated individually with the Johnson/Cousins 
$V$ and $I_C$ using an offset between the instrumental magnitude and the calibrated one. 
Calibrated magnitudes are averaged separately
within two groups, short ($< 100$ sec) and long ($\ge 200$ sec) exposures. 
Comparison of the short- and long-exposure 
calibrated magnitudes yields differences of $\sim 0.03$ mag in $V$, and $\sim 0.01$ in $I$; we regard these offsets as 
estimates of the limitations of our calibration. Individual photometric errors 
obtained from the scatter of measurements for any given object are $\sigma_V = 0.01 $ between $V=16.8-20.2$ and $\sigma_I = 0.008$ between $I_C=17-21$. 
These we consider to be well-measured objects, with errors increasing gradually toward fainter magnitudes.  
The short- and long-exposure magnitudes are then combined,
discarding saturated measurements from the long exposure and poor signal-to-noise measurements from the short exposures. 
The $V,I_C$ data of the early- and intermediate-epoch data reach faint limits of the order of $V\sim 24$ and $I_C\sim 23.5$ (defined by
the turnover in the object counts in each band). Our final proper-motion catalog, however, is shallower due to the 2012 data.
We use our calibrated magnitudes to select objects in broad ranges of $I$ magnitudes 
and $V-I_C$ colors and to explore magnitude- and color-dependent systematics in the astrometry.

\section{Object Classification}
For the task of object classification, 
we use only $I_C$-band data of the intermediate epoch as these have the best FWHM and area coverage 
(see Figs.~\ref{fig:fig1} and ~\ref{fig:fig3}). From these, we discard all observations with FWHM$> 1\arcsec$.
SExtractor's neural-network classifier based on the input FWHM is used to determine a star/galaxy class parameter. The  average FWHM 
estimate for each chip and exposure is provided by the SDFRED1/2 pipeline.
The class parameter is used in conjunction with the ratio of the semimajor to semiminor axis, $a/b$, also given by SExtractor.
For each object, average values for these parameters are calculated over the number of observations.
We employ the same methodology developed and tested in our previous work \citep{cg16} 
to select unblended stars and galaxies with shapes somewhat similar to stars.
This methodology aims to provide clean samples of stars and galaxies rather
than complete samples; thus, a large portion of objects will be marked as ``unclassified''.

\begin{figure}
\includegraphics[width=\columnwidth,angle=-90]{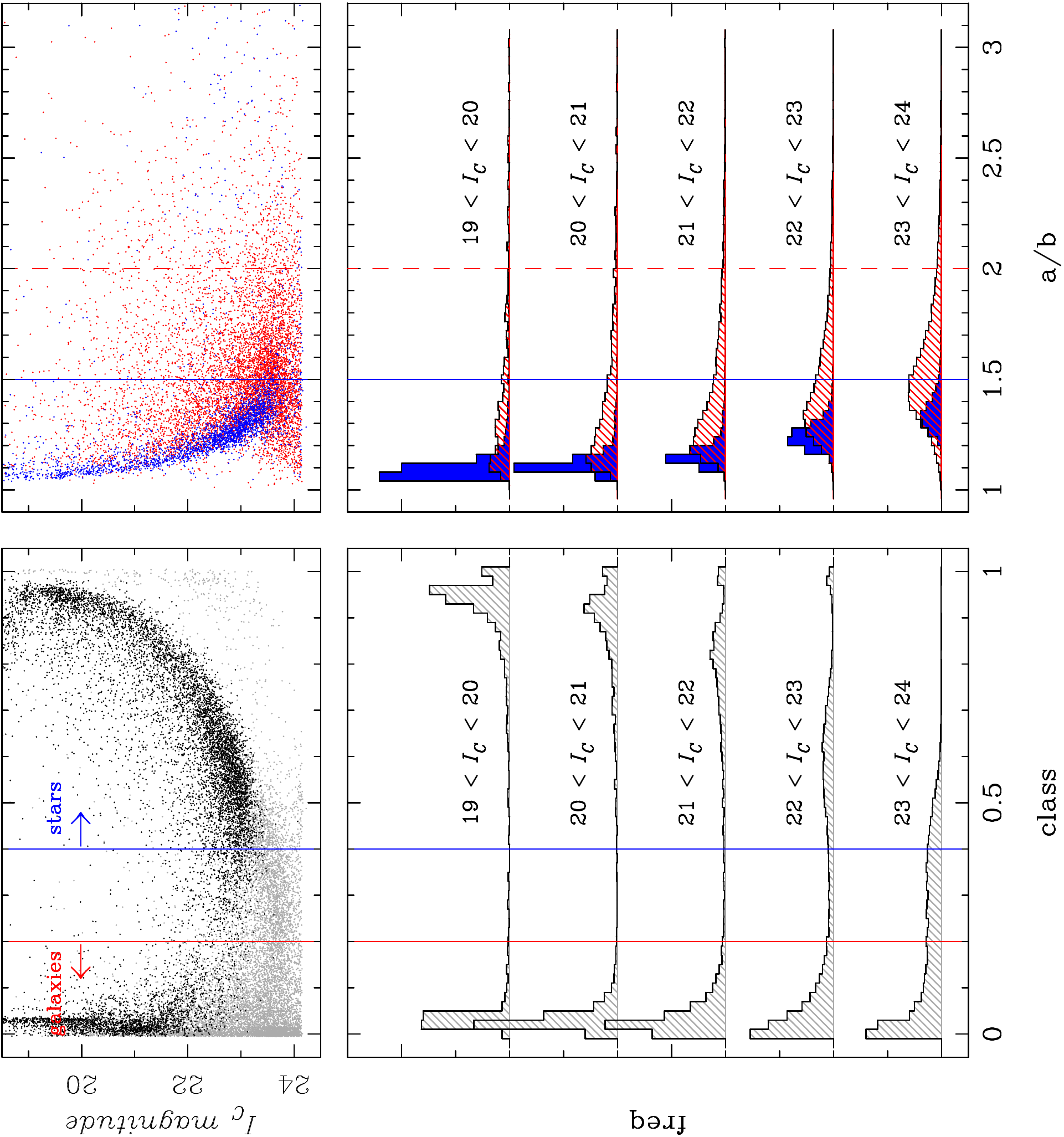}
\caption{Object magnitude versus class parameter (top left) and versus axis ratio $a/b$ (top right).
Grey points in the top left panel show $20\%$ of our entire sample as classified from the 2005/6 $I_C$-band data in the area of 
the proper-motion study. 
Black points show objects with proper-motion measurements (a shallower sample due to the 2012 observations). Histograms show
the distribution of class and $a/b$ parameters for different magnitude bins. 
The vertical lines show our cuts for selecting stars and galaxies.}
\label{fig:fig4}
\end{figure}

In a first cut, we choose as stars objects with class $\ge 0.4$, 
and as galaxies objects with class $\le$ 0.2. Next, we further limit the galaxies to be 
those with $a/b \le 2.0$; and stars, those with $a/b \le 1.5$. 
In Figure~\ref{fig:fig4} we show $I_C$ magnitude as a function of average object parameter: 
class (top left panel) and $a/b$ (top right panel). Distributions of the class (left)  and $a/b$ (right) parameters for
different magnitude bins are shown in the lower panels. The two peaks in these distributions, corresponding to 
stars and galaxies, are well separated down to $I_C \sim 23$. In the top left panel of Fig.~\ref{fig:fig4}, we also highlight 
the subsample of objects with measured proper motions; this is substantially shallower than the sample with object classification
due to the 2012 observations being taken in a narrower-band filter compared to the 2005/6 observations. 
Thus, contamination issues of star/galaxy samples should not be an issue in the final proper-motion measurement of Sextans.

\section{Astrometry}

\subsection{Distortion-Correction Maps}

We model the Suprime-Cam detector at each epoch by stacking residuals and constructing distortion maps in one of two 
ways: 1) internally, using data sets with many offsets and dithers or 2) externally, using 
an astrometric catalog close to the epoch of observation. Modeling at each epoch is required
since the camera underwent a major upgrade in 2008, including the installation of new detectors. In addition, the
2012 data were taken with intermediate band Str\"{o}mgren $b$ and $y$ filters, while the early and intermediate-epoch
data were taken with wide-band Johnson filters. In principle, the 2002/3 and 2005/6 data sets should not require
separate maps, as the same detector is used. Nevertheless, we construct separate maps to check for variations, and to
account for possible changes in the optics and detector with time.   

We apply the first method to the 2002/3 and 2005/6 data, and the second
method to the 2012 data. Specifically, for the 2012 data, we use the $Gaia$ DR1 catalog \citet{gaia16} which is ideal 
for this purpose given its astrometric properties and its near-contemporary 2015 epoch. 
The Sextans-field data, however, were insufficient for the entire calibration process as the 2002/3 exposures did not
provide enough offsets/dithers to build a reliable map.  Adding to our difficulties, $Gaia$ DR1 contains an unfortunate
gap at this position in the sky, due to the satellite's scanning law.
Fortunately, the SMOKA archive includes a set of observations in the field of NGC 2419 very similar to that of Sextans.
Specifically, we found $V$ and $I_C$ observations taken in 2002 and 2005, and Str\"{o}mgren $b$ and $y$ taken in 2012.
The 2002 NGC 2419 field included sufficient offsets/dithers for reliable mapping, and the $Gaia$ DR1 data are 
excellent in this area. In addition, the NGC 2419 field is at a lower Galactic latitude than Sextans, thus providing 
many more stars to build the residual maps.

\begin{figure}
\includegraphics[width=\columnwidth,angle=0]{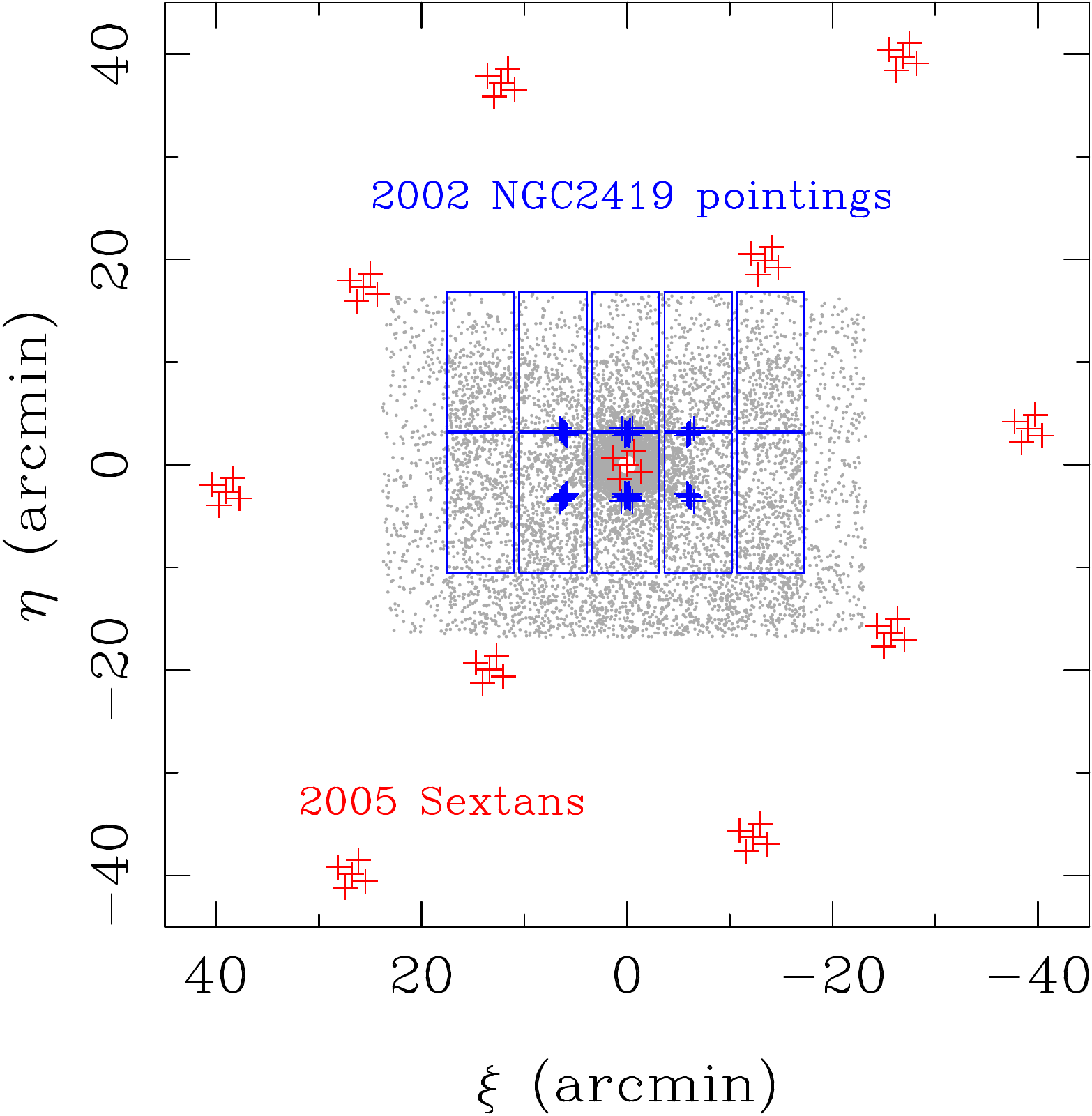}
\caption{Offsets and dithers of the observations used to build distortion maps in the field of NGC 2419 (2002 epoch) and in the field of Sextans (2005/6 epoch).The grey dots represent the proper-motion catalog in the field of NGC 2419.
}
\label{fig:fig5}
\end{figure}

In Table~\ref{tab:tab2} we summarize the characteristics of the constructed distortion-correction maps; $N_d$ represents
the total number of offsets/dithers available. The number of exposures is larger than this, since some are repeats at 
different exposure times and filters. 
\begin{table}
\caption{Distortion Map Properties}
\label{tab:tab2}
\begin{tabular}{lllll}
\hline
\multicolumn{1}{c}{Epoch} & \multicolumn{1}{c}{Field} & \multicolumn{1}{c}{Filter} & \multicolumn{1}{c}{$N_{d}$} & \multicolumn{1}{c}{Method} \\
\hline
\multicolumn{1}{l}{2002} &\multicolumn{1}{l}{NGC 2419}  & \multicolumn{1}{l}{$V$} & \multicolumn{1}{r}{44} & \multicolumn{1}{r}{internally} \\
\multicolumn{1}{l}{2005/6} &\multicolumn{1}{l}{Sextans}  & \multicolumn{1}{l}{$I_C$} & \multicolumn{1}{r}{88} & \multicolumn{1}{r}{internally} \\
\multicolumn{1}{l}{2012} &\multicolumn{1}{l}{NGC 2419}  & \multicolumn{1}{l}{$b$,~$y$} & \multicolumn{1}{r}{5} & \multicolumn{1}{r}{Gaia DR1} \\
\hline
\end{tabular}
\end{table}
In Figure~\ref{fig:fig5} we show the pattern of offsets and dithers for the 2002-epoch data (NGC 2419 field) and for the 2005/6
epoch (Sextans field).
The origin of the plot represents the center of the each target field (NGC 2419/Sextans). 

The distortion maps are constructed using the following steps:

1) Transform the $X,Y$ system of each CCD chip onto a common system. For epochs 2002/3 and 2005/6, we use the SDSS
as reference. For the 2012 epoch,
we use $Gaia$ DR1 in the field of NGC 2419 and SDSS in the field of Sextans. Polynomials up to third order are employed.
In the field of Sextans, between 100 and 270 stars define these transformations. In the field of NGC 2419, between 160 and 280 
stars are used, except for the chips that include the cluster, where several hundred stars are used.
Next, we combine all equatorial position measurements at a given epoch to form an average working catalog, 
retaining only those objects with a minimum number of repeated measurements; three in fields/epochs with fewer exposures, 
but generally five.

2) At each epoch, perform a new transformation of each chip and exposure into the average catalog built in step 1),
using up to third-order polynomials. This second transformation eliminates the uncertainty introduced by the proper-motion
scatter due to the epoch difference between Subaru observations and the original reference catalog chosen. It also 
uses many more objects (between a few hundred to a few thousand) per transformation as the Subaru exposures are 
deeper than either $Gaia$ DR1 or SDSS. More importantly, by averaging the positions of an object as it falls on 
different chips or in different regions of the same
chip, position-dependent systematics can be alleviated. 
The extent to which the averaging will reduce such systematics depends on the number and form of the
dither/offset patterns in the available archive observations. With this in mind, we have chosen the Sextans 2005/6 dataset
to construct a distortion-correction map at this epoch, based on residuals from transformations of these data into the
average positional catalog.
This map represents distortions beyond those of the third-order polynomial used in the transformations that formed it, 
and can be applied to both Sextans and NGC 2419 data sets at 2005/6. 
For the 2002 epoch, we use the NGC 2419 data set in a similar manner to build a 2002-epoch map. 
Finally, for the 2012 data set, we use 
the residuals obtained directly from the transformation of Subaru data into the $Gaia$ DR1 positions. Here, the need
for many dithers is not as critical, as the $Gaia$ DR1 positions are at a similar epoch to the Subaru observations, and are 
much more accurate and precise than the Subaru positions. Specifically, the standard errors of the transformations of
Subaru 2012 positions into $Gaia$ DR1 are $\sim 6$ mas for well measured objects. The typical positional precision of Subaru
data, as determined from repeats at the same epoch, is of the order of $5 - 6$ mas \citep[see also][]{cg16}, thus showing that
Subaru errors dominate the error budget. Estimated positional 
uncertainties in the $Gaia$ DR1 NGC-2419 field are of the order of 0.1 mas, increasing with magnitude, such that at
$G=19.8$ the error is 1 mas. This is consistent with the standard errors of our transformations. 
Therefore, residuals from the transformations of
Subaru into $Gaia$ DR1 are used directly to construct the 2012 map.

3) Construct maps based on stacked residuals falling within spatial cells that are at a lower resolution than the Subaru chips.  
The maps consist of an array of cells, each composed of 32x32 Subaru CCD pixels
for the 2002 and 2005/6 data, and 64x64 CCD pixels per cell for the 2012 data. 
Of the order of 100 residuals are thus accumulated
in each map cell for the early data, while for the 2012 maps the average number is closer to 20.
An oversampling radius of twice the size of the cell is used to calculate the average residual in each cell. 
Next we apply an `\`{a} trous' wavelet filtering \citep{sm02} to further smooth the somewhat noisy maps without
adversely broadening their features. Once initial maps are constructed
we apply them to the original pixel coordinates and repeat steps 1) and 2). 

Inspecting the newly formed map of residuals, it is apparent that
the first application of the correction map actually undercorrects, most likely due to the oversampling smoothing. 
For this reason, we choose to use the correction maps with a multiplicative factor.
To derive the best value for this factor, we make a bracketed ``guess'' of 2 for its value and use this to correct the
original positions and generate yet another map following the same procedure as above.
As expected, inspection of the new post-correction map shows that, this time, we have overcorrected. 
However, from the residuals derived from the two different applications of the map -- one with a factor of 1 and the other 
with a factor of 2 -- it should be possible to find the optimal value for the factor.
We do so by solving for the linear combination of factor-1 residuals with factor-2 residuals that yields zero for the
combined residuals.
Doing so for the various maps (different chips and epochs and in two coordinates), 
produces a range of values for the factor, with a mean near 1.2 but with a scatter of about 0.3.
A test run using the single value of 1.2 produced post-correction maps that still showed structure in the residuals.
Thus, a simple scalar value appears inadequate.

Instead, a ``map'' of factor values can be generated for each chip, again using the factor-1 and factor-2 residuals, but
finding the linear combination that yields zero summed residuals on a cell-by-cell basis.
Note that in the interpolation used to generate these factor maps there is a division by the difference in factor-1 and factor-2
cell values.
Thus, in a cell where these are similar to one another, the interpolated value is poorly determined.  
For this reason, the factor maps themselves are value-clipped and smoothed before being multiplied by the corresponding
original distortion-correction maps.
The final result is a set of calibrated distortion-correction maps for each Subaru chip, at each epoch, and along both
CCD directions.

\begin{figure*}
\includegraphics[scale=0.70]{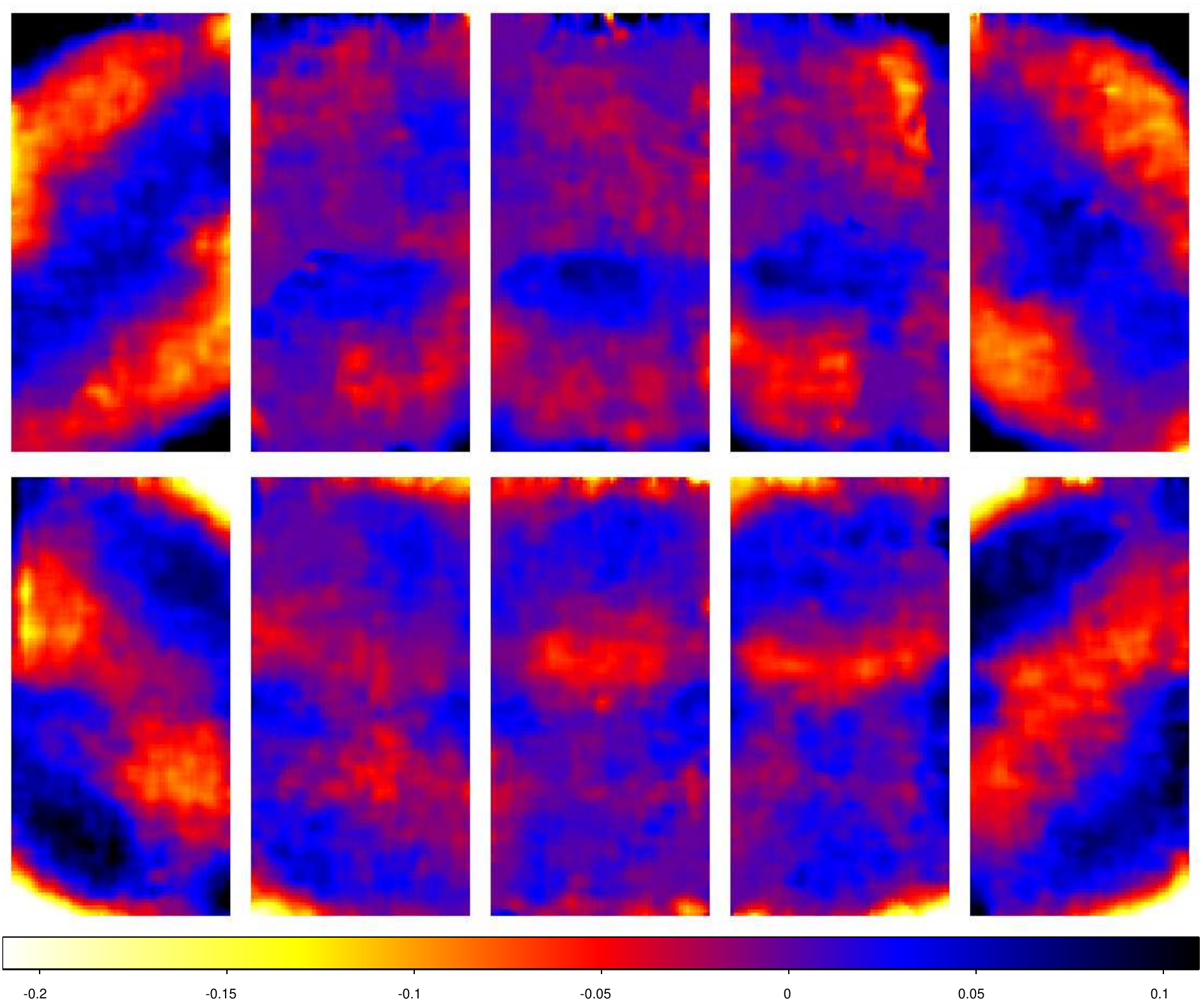}
\includegraphics[scale=0.70]{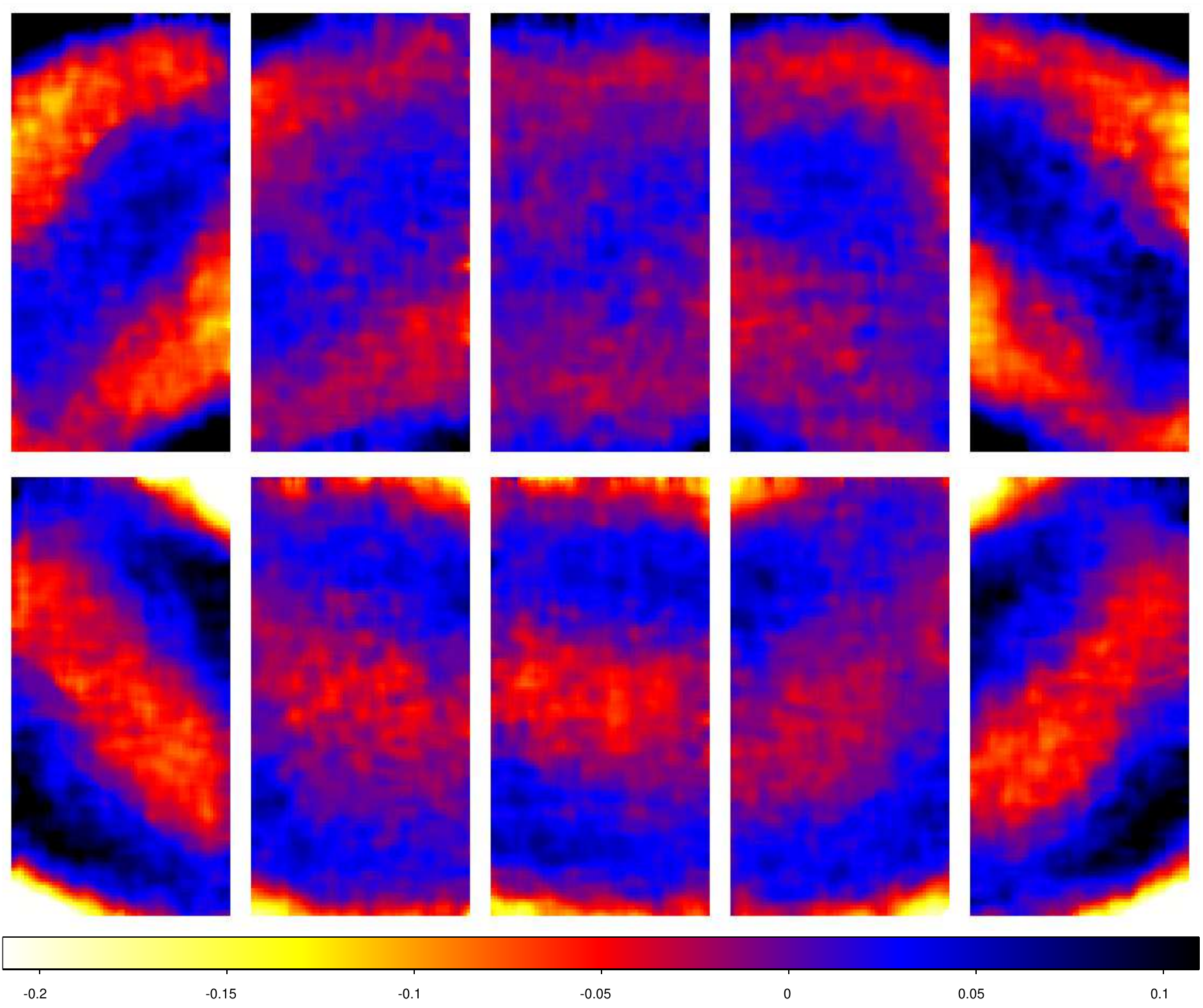}
\caption{Distortion-correction maps beyond third-order field terms for Suprime-Cam in 2002 (top) and 2005 (bottom).
Corrections in $Y$-coordinate are shown; those in $X$ coordinate are similar. 
Map resolution is $32 \times 32$ Suprime-Cam pixels.
Z-scale units are Suprime-Cam pixels with a scale of $0\farcs2$/pix.}
\label{fig:fig6}
\end{figure*}

\begin{figure*}
\includegraphics[scale=0.70]{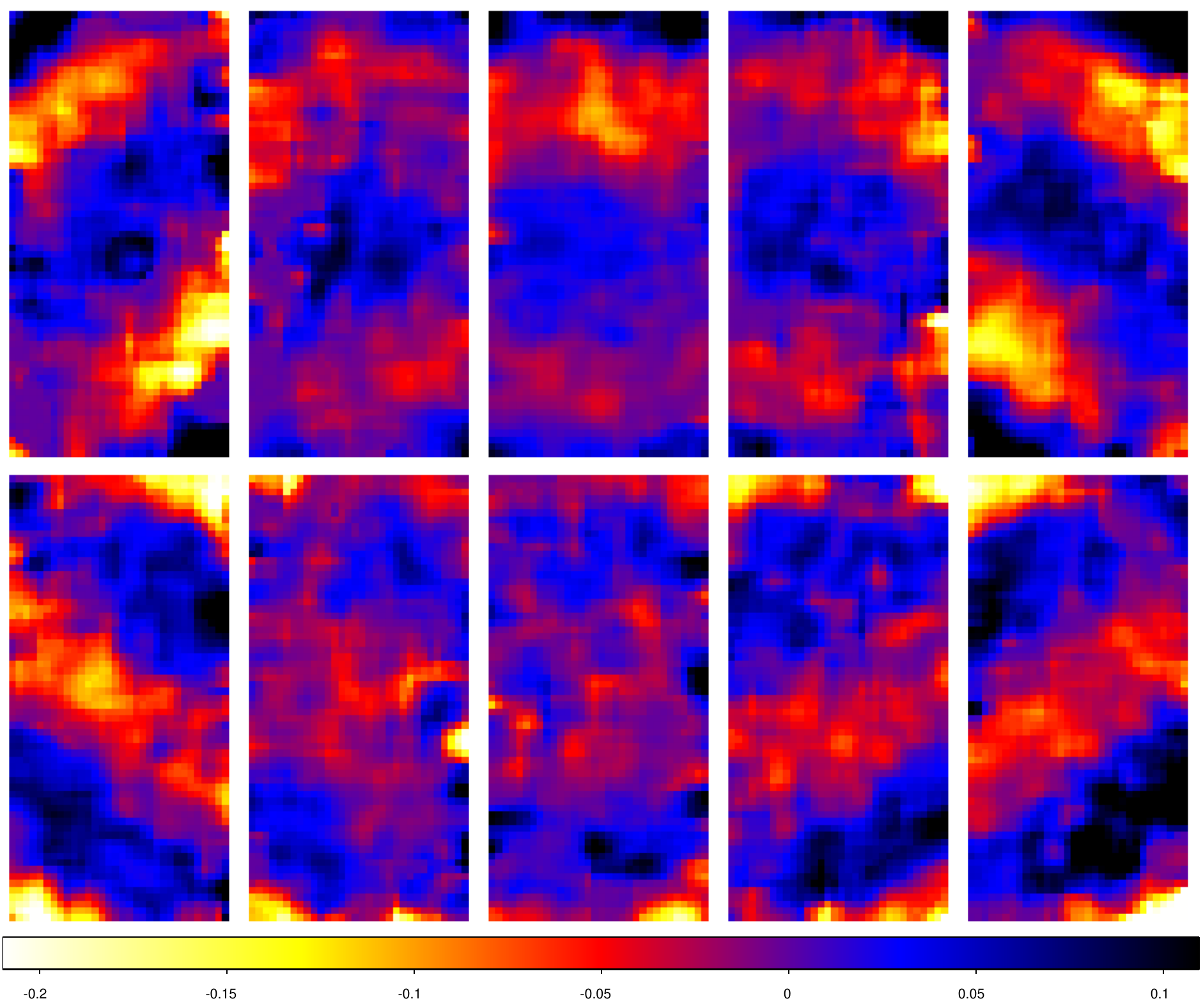}
\caption{As in Fig.\ref{fig:fig6}, but for the 2012 data. Map resolution is $64 \times 64$ Suprime-Cam pixels.
%Distortion-correction maps beyond third-order field terms for Suprime-Cam in 20%02 (top), 2005 (middle) and 2012 (bottom).
%Corrections in $Y$-coordinate are shown; those in $X$ coordinate are similar. U%nits are pixels with a scale of $0\farcs2$/pix.
}
\label{fig:fig7}
\end{figure*}

In Figures~\ref{fig:fig6} and \ref{fig:fig7} we show the final distortion-correction maps in $Y$-coordinate at each epoch.

These final correction maps were checked by analyzing the post-correction residuals one more time and verifying that the
residuals were spatially flat and had a mean value of zero, although there was a slight deviation seen in the outermost corners 
of the outermost chips.
Overall, the rms of the post-correction residuals was lowered by use of the final correction maps.
For instance, for the 2005 epoch data, the mean (over ten chips) standard deviation of the residuals decreased from 0.010 pixels
using the factor-1 correction map down to 0.006 pixels using the final correction map.

The 2005/6 data were rich enough that two sets of maps could be constructed, for the $I_C$-band data 
and for the $V$-band data, separately. 
No significant differences in the two sets of maps were found. We will assume that this also holds for the 2002/3 data. 
Therefore, the 2002 $V$-maps in the field of NGC 2419 are applied to both $V$ and $I_C$ data of Sextans in 2002 and 2003. 
The 2005/6 $I_C$-band maps are applied to all of the 2005/6 data. 
Finally, the 2012 $b$ and $y$ maps, made from observations in the field on NGC 2419, are applied
to both Sextans and NGC 2419 data.  Equatorial positions thus corrected are transformed via a gnomonic projection into 
standard coordinates $(\xi,\eta)$ with the origin at Sextans' center. These are the positions used
in the proper-motion determination.

\subsection{Global Solution: Relative Proper Motions}
\begin{figure}
\includegraphics[width=\columnwidth,angle=0.]{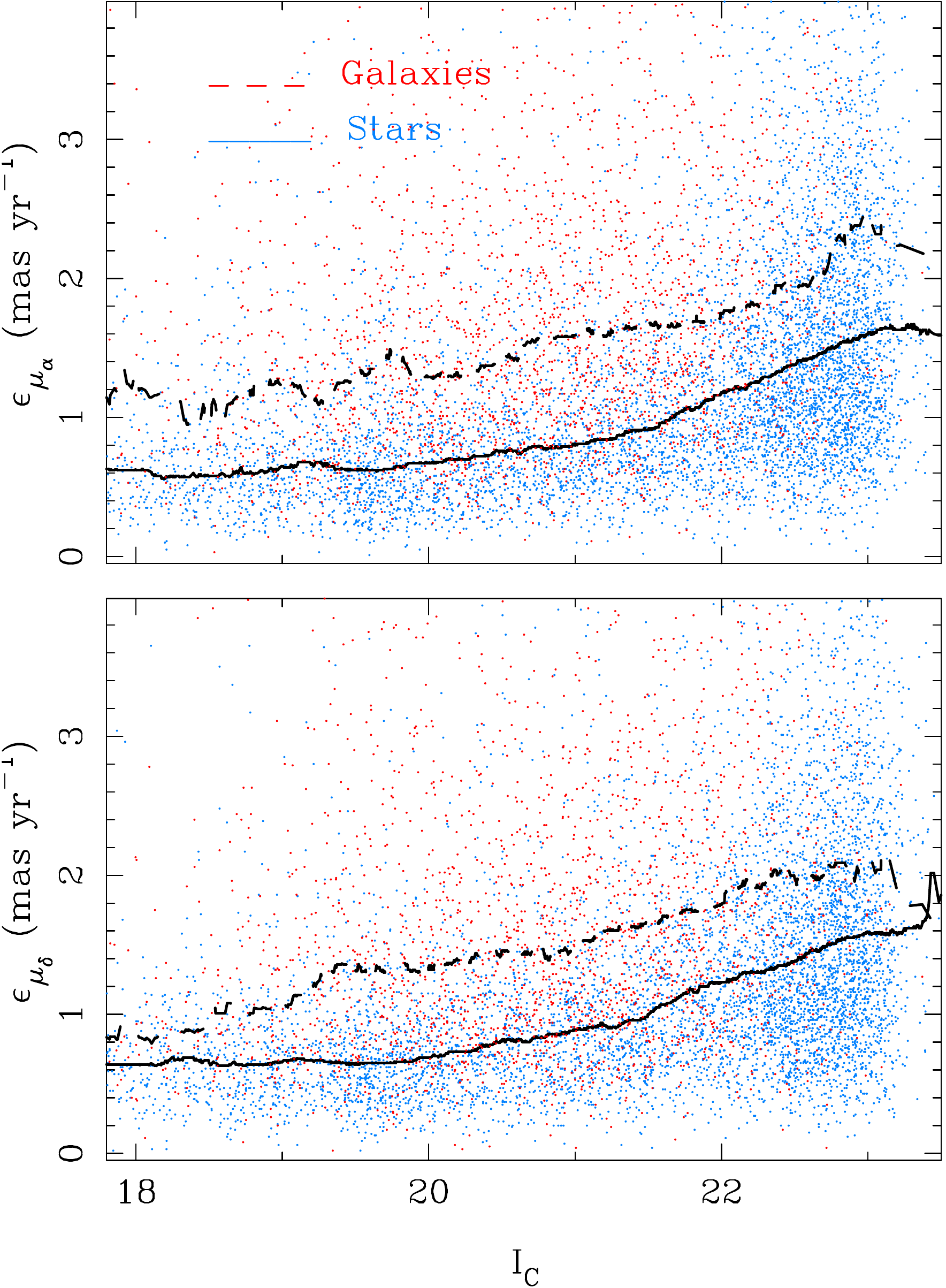}
\caption{Proper-motion uncertainties as a function of $I_C$ magnitude for stars (blue) and galaxies (red). Moving
medians for galaxies (dashed line) and stars (continuous line) are also shown.
}
\label{fig:fig8}
\end{figure}

Proper motions are based on a total of 2460 individual frames spanning the three epochs. 
To begin, we assemble these frames into a master list
with unique object identifier using a matching tolerance of $0\farcs5$. 
A plate overlap solution \citep[e.g.,][]{eich88} is 
used to determine each object's position and proper motion as well as the ''plate'' constants 
\citep[see also our detailed description in][]{cg16}. 
Considering the large number of frames involved, it was decided to work with
average catalogs per epoch as opposed to a simultaneous solution, in order to more easily identify and discard poor frames.
These per-epoch catalogs are made from map-corrected (see Section 6.1) positions of 
individual frames transformed into 
each epoch's initial average catalog (i.e., step 1 in Section 6.1). We create six such average ``plates''  for: 
fall 2002, spring 2003, February 2005, May 2005, December 2005 - January 2006, and 2012 (see Tab.~\ref{tab:tab1}), 
and use these in a
plate-overlap solution. The average number of measurements per object over the most
relevant magnitude range ($I_C \sim 19 - 22$, see Section 6.3) at each of these epochs is 4, 7, 6, 16, 9, and 9, respectively.
Transformation between the six average plates employs up to fourth-order polynomial terms, and both stars and galaxies are used as
 reference objects. Mean-epoch positions and relative proper motions are determined as unweighted least-squares fitting 
of $\xi$ and $\eta$ standard coordinates as linear functions of time. 
The scatter about the linear fits gives individual proper-motion uncertainty estimates.
The procedure is applied iteratively with convergence after ten iterations.
Formal proper-motion uncertainties as a function of magnitude are shown in Figure~\ref{fig:fig8} for stars and galaxies. Moving
medians are also shown as these are a better statistical representation of the true errors. As is typical, galaxies have
formal uncertainties about a factor of two larger than those of stars.
For $99\%$ of the objects with calculated proper motions the time span is between 9.7 and 10.1 years. The remaining $1\%$
have time spans between 7.6 and 7.8 years.    

\subsection{Local Solution: Absolute Proper Motions}
In theory, the difference between the mean relative proper motion of Sextans members and that of the sample of
reference galaxies will yield the absolute systemic motion of Sextans.
Unfortunately, such a simple procedure is
not adequate for two main reasons. First, the reference system of the relative proper-motion solution consists of a mixture
of all types of objects -- galaxies, Sextans stars, and Galactic foreground field stars -- all of which are 
distinct populations with different kinematical properties. 
Combined with the possible non-uniform spatial distributions for these populations, this can introduce artificial gradients 
over the area studied.
Second, it is possible that flaws remain in our distortion maps due to the limited number of residuals available to map
the entire area of each detector, which could result in geometric systematics in the relative proper motions.
On a sufficiently small scale length, i.e., locally, we can assume that such systematics disappear if we refer each Sextans 
star to a group of its neighbouring galaxies. 
In practice, we refer each galaxy to a local reference system of nearby Sextans stars, as the stars are better measured allowing
to measure each galaxy offset best, and therefore better define the remaining geometric systematics. 
In addition, this study is shallower than our \citet{cg16} study in the field of the Draco dwarf spheroidal galaxy, 
thus providing a less dense network of galaxies and a less well-determined local solution, if directly based on galaxies.
Lastly, Sextans has such a large spatial extent, that our wide-field study is still located within its core radius 
(see Fig.\ref{fig:fig1}), providing an abundant number of Sextans stars. 
In what follows, the mean proper motions of various samples are calculated with respect to nearby Sextans members.
Simple inversion of the proper motion of galaxies with respect to Sextans stars will yield 
the motion of the dwarf with respect to the inertial reference system of galaxies.

\subsubsection{Local Solution: Sextans Samples}
\begin{figure}
\includegraphics[width=\columnwidth,angle=0.]{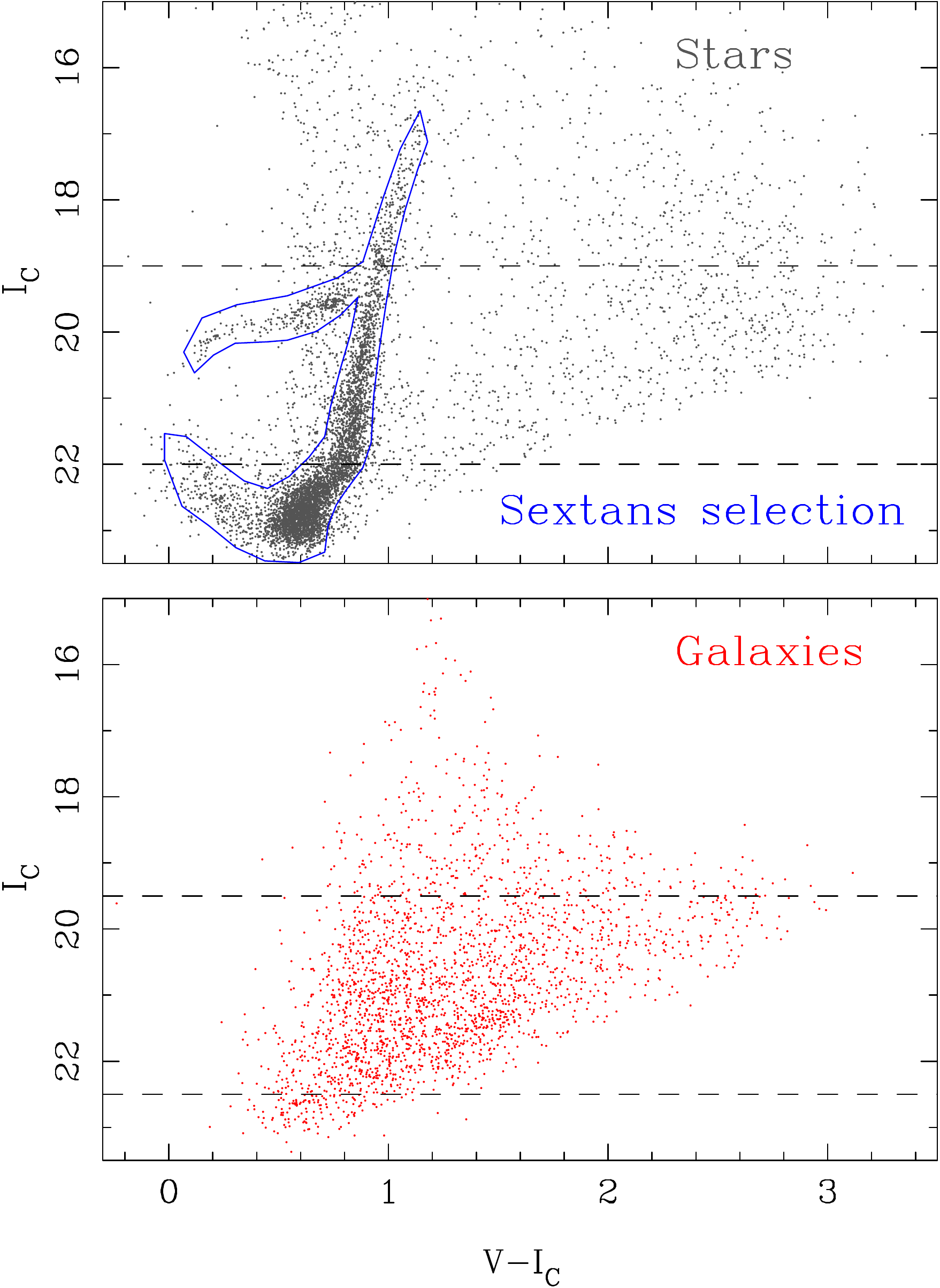}
\caption{
Color-magnitude diagram in the field of Sextans. Stars are shown in the top panel, together with the
selection of Sextans members, while galaxies are shown in the bottom panel. Dashed lines indicate
the magnitude range for Sextans stars and galaxies to be used in the final proper-motion determination. 
}
\label{fig:fig9}
\end{figure}

Sextans members are selected photometrically, from the color-magnitude diagram (CMD) based on our $VI_C$ photometry 
(see Section 4). In Figure~\ref{fig:fig9} we present this selection; the Sextans CMD sequence is outlined by the dark contour. 
Besides our CMD-selected sample, we also form another 
Sextans sample based on the more accurate photometry from \citet{lee03}, in a similar manner as in Fig. ~\ref{fig:fig9}.
Note that the \citet{lee03} photometry did not fully cover our study area.
We also identify a sample of horizontal-branch members to check for color trends. These are selected using our photometry.
Finally, a sample of radial-velocity (RV) members from the studies by \citet{wal09a} and \citet{wal09b}, stars having RV 
memberships $\ge 70\%$, are also selected to check magnitude-dependent trends. 
This latter sample includes predominantly red giants; thus it is at the bright end of our study.
These other Sextans samples are constructed in order to check on our method for absolute proper-motion determination,
verifying that magnitude-dependent effects, for example, are under control.

Dashed lines in Fig.~\ref{fig:fig9} show the bright and faint limits used in our study for Sextans stars and for galaxies.
Brighter than $I_C = 19.0$ for stars and 18.5 for galaxies, systematics related to saturation affect the proper motions.
Fainter than $I_C = 22.0$ for stars, the increase in photometric errors contribute to contaminating the 
Sextans sample with field stars, such than the mean motion is biased at levels approaching our final uncertainty estimate. 
For galaxies, this limit is $I_C = 22.5$, beyond which proper-motion errors for galaxies are large, and the star/galaxy 
separation is less accurate (see Section 5).

The local solution is done as follows. For each target object (galaxy or star) we choose the closest $N_n$ reference objects 
(Sextans stars) and determine their median proper motion separately in each coordinate. This median is to be subtracted directly from the proper-motion of the 
target object.  In some cases the target is not well centered within the surrounding reference objects. If so, an additional 
correction to this median value is made based on the gradient of the reference objects' proper motions as a function of
$\xi$ and $\eta$ and the spatial offset between the target's position and that of the mean of the reference objects.
Once every target's proper motion with respect to the local reference system is determined, we calculate 
the mean for the ensemble of targets as follows. We restrict the sample spatially to within $-800\arcsec \le \xi \le +800\arcsec$,
and $-700\arcsec \le \eta \le +700\arcsec$ since the edges of the field are less well-modeled.  Also, targets with
proper-motion offsets larger than 10 mas~yr$^{-1}$ are discarded. 
Doing so, we determine the proper-motion mean and the statistical uncertainty in the mean
from probability-plot estimates based on the inner $80\%$ of the distribution in each coordinate \citet{ham78}. 

We demonstrate the precision of our local-solution method by applying it to the various Sextans samples, each of which
should yield a mean offset motion of zero.
First, the sensitivity of the method to the scale length of the local solution is explored by using our photometrically
selected Sextans sample as both reference and target groups and varying the number of reference stars used to calculate the
local adjustments. 
Then, keeping as reference our photometric sample, we substitute for the target group our variously selected Sextans subsets:
horizontal-branch (HB) members, CMD-members from the photometry of \citet{lee03}, and finally, 
RV members from \citet{wal09b}.  In Table~\ref{tab:tab3} we present the results of these tests. 
Keep in mind the reference system consists of the
Sextans CMD-selected members within $I_C=19 - 22$ for all tests. The ensemble of target objects is described in the first column of 
Tab.~\ref{tab:tab3}. Mean proper-motion estimates in each coordinate are listed in columns 2 and 3. The total number of 
target stars used in each estimate $N_t$ is shown in column 4, and in columns 5 and 6 we show the number of 
reference stars used in the local system $N_n$ and the average radius of the local system, respectively.

The first five entries in Tab.~\ref{tab:tab3} demonstrate variation of the size of the local reference system.
Statistically, only the solution with $N_n=200$ is significantly different from zero in $\mu_{\alpha}$. However, 
formal uncertainties decrease with the size of the local system. For this reason we adopt the solution with 
$N_n=20$ as our preferred scale-length solution, although it is comparable to the $N_n=50$ or $N_n=10$ solutions.
The subsequent entries in Tab.~\ref{tab:tab3} explore different selections of Sextans members as target groups, 
all with the local-system parameter set to $N_n=20$. 
The solutions for all four targets, with $N_n=20$, are consistent with one another within their formal uncertainties, 
and consistent with a value of zero. 
This is an indication that systematics due to sample contamination with field stars or due to magnitude and/or color, if present, 
are below the level of our formal uncertainties. 
Indeed, averaging the four solutions of different Sextans samples and with $N_n=20$, we obtain 
averages consistent with zero with standard deviations $\sigma_{\mu_{\alpha}} = 0.034 $  mas~yr$^{-1}$, and
$\sigma_{\mu_{\delta}} = 0.046 $ mas~yr$^{-1}$. We will incorporate these values into the total error budget of Sextans' proper 
motion, as they are indicative of limitations due to likely leftover systematics in the Sextans reference sample. 
\begin{table*}
\caption{Testing the local solution: various Sextans-star samples referred to CMD-selected members with $I_C = 19-22$.}
\label{tab:tab3}
\begin{tabular}{lrrrrr}
\hline
\multicolumn{1}{c}{Sample} & \multicolumn{1}{c}{$\mu_{\alpha}$} & \multicolumn{1}{c}{$\mu_{\delta}$} & \multicolumn{1}{c}{$N_t$}  
& \multicolumn{1}{c}{$N_{n}$} & \multicolumn{1}{c}{$r_{local}$} \\
& \multicolumn{1}{c}{(mas~yr$^{-1}$)} & \multicolumn{1}{c}{(mas~yr$^{-1}$)} & & & (arcsec) \\
\hline
Sext. CMD $I_C=19-22$ & $-0.084\pm0.041$ & $-0.001\pm0.045$ & 1331  & 200 & 212 \\
Sext. CMD $I_C=19-22$ & $-0.058\pm0.037$ & $-0.003\pm0.042$ & 1331  & 100 & 150 \\
Sext. CMD $I_C=19-22$ & $-0.028\pm0.034$ & $-0.031\pm0.039$ & 1332  &  50 & 104 \\
Sext. CMD $I_C=19-22$ & $-0.023\pm0.034$ & $-0.053\pm0.036$ & 1332  &  20 &  63 \\
Sext. CMD $I_C=19-22$ & $-0.020\pm0.030$ & $-0.048\pm0.034$ & 1333  &  10 &  41 \\
\hline
Sext. HB              & $ 0.021\pm0.044$ & $ 0.044\pm0.048$ &  210  &  20 &  64 \\
Lee03 CMD $I_C=19-22$& $-0.002\pm0.034$ & $-0.046\pm0.038$ & 1138  &  20 &  63 \\
Walker09 RV $I_C=19-22$& $-0.051\pm0.066$ & $-0.044\pm0.061$ & 111  &  20 &  61 \\
\hline
\end{tabular}
\end{table*}

\begin{figure}
\includegraphics[width=\columnwidth,angle=0.]{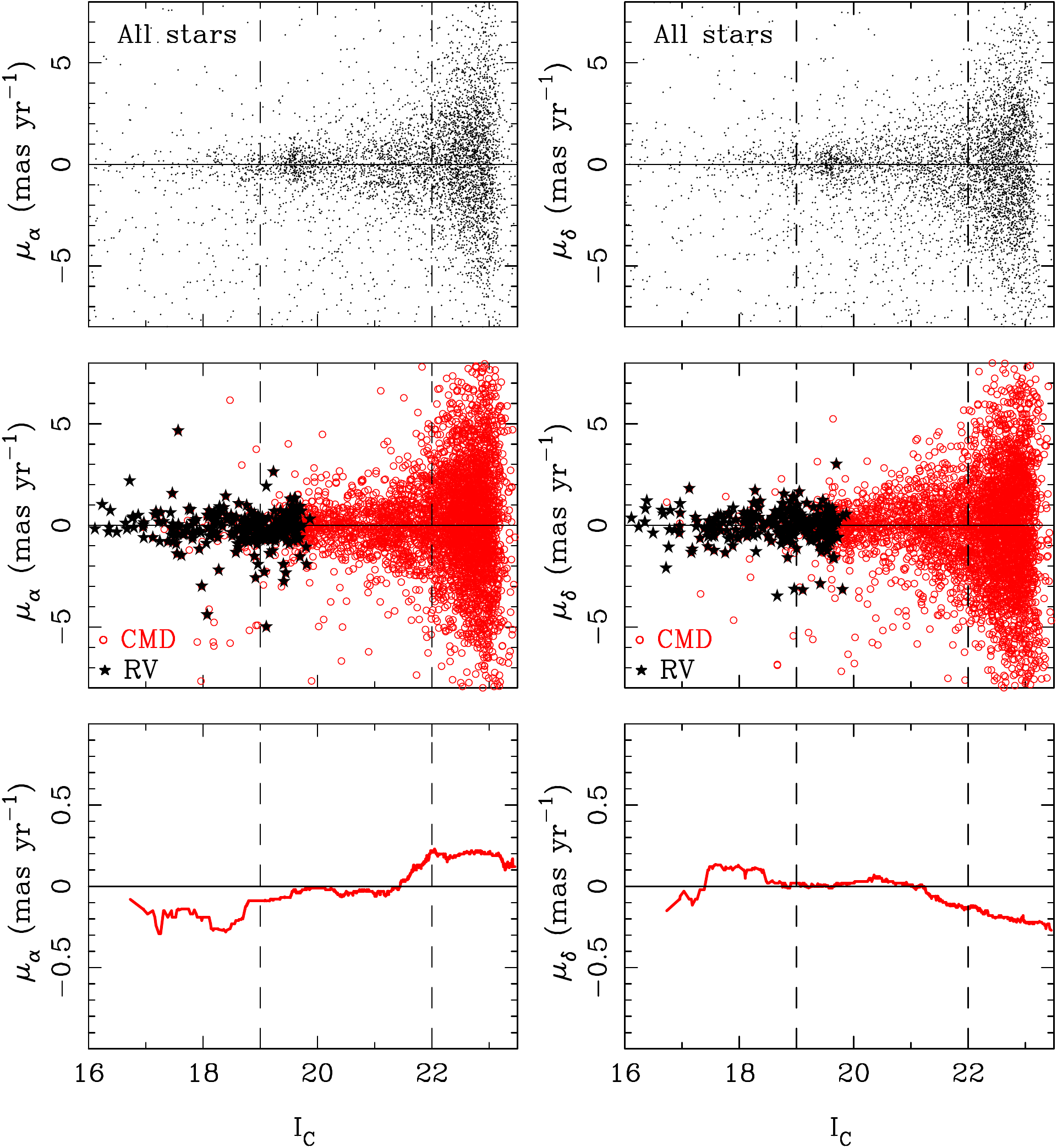}
\caption{
Local-solution proper motions of stars as a function of magnitude. The local reference system consists of Sextans stars (see text).
The top plots show all objects classified as stars; middle plots show Sextans members  
based on the CMD photometric selection (round/red symbols) and on the
radial-velocity selection (star/black symbols). For Sextans members, the locally-corrected proper motions should be zero.
The bottom plots show moving medians of proper motions with magnitude
for the CMD-selected Sextans members. Note the zoomed-in proper-motion scale compared to the previous plots. 
Saturation affects the bright end, while foreground field
contamination affects the faint end. Dashed vertical lines indicate magnitude limits adopted for the final reference sample of
Sextans stars.
}
\label{fig:fig10}
\end{figure}

Exploring possible magnitude-related effects more directly,
in Figure~\ref{fig:fig10} we show the locally--corrected proper motions of stars as a function of magnitude.
Again, since the reference frame 
consists of CMD-selected Sextans members (with $I_C = 19 -22$), proper motions should be zero for Sextans members.
The top panels show all objects classified as stars, the
middle panels show Sextans members selected from the CMD and from radial velocities, and the bottom panels
show moving medians with magnitude of the CMD-selected Sextans members. 
The vertical dashed lines indicate our imposed magnitude limits that were, in fact, chosen based on the bottom panels, i.e.,
the magnitude range over which deviations are minimal.
Proper motions at the bright end are affected by saturation, while at the faint end foreground field-star contamination
likely biases the photometric samples.
Also, fainter than $I_C=22$, the proper-motion scatter increases substantially due to increased proper-motion errors.

\begin{figure}
\includegraphics[scale=0.33,angle=-90.]{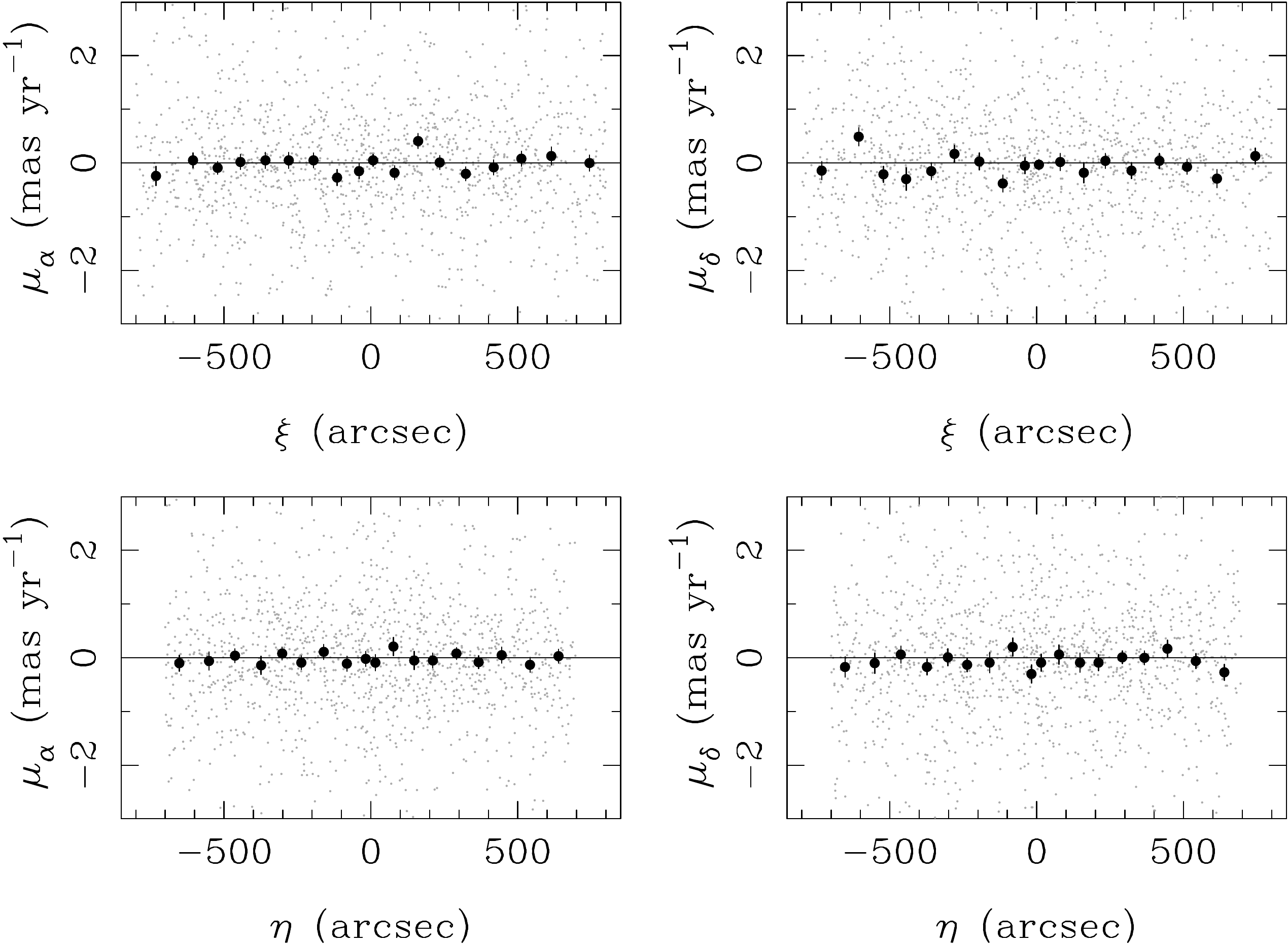}
\caption{
Local-solution proper motions of Sextans stars ($I_C=19-22$) as a function of position over the field of study (grey 
symbols).
The local reference system is also made up of Sextans stars (see text), thus the mean should be zero.
Averages in equal-number-of-stars bins as a function of the $\xi$ and $\eta$ positions are also shown.  
No significant position trends are apparent.}
\label{fig:fig11}
\end{figure}

Inspection of local-solution proper motion of Sextans members within $I_C =19 - 22$ as a function of 
position in the field shows no discernible trends. Figure~\ref{fig:fig11} shows these proper motions as a function of 
standard coordinates $\xi$ and
$\eta$ for individual Sextans stars, and means over bins composed of equal numbers of stars. 

\subsubsection{Local Solution: Galaxies}

Local-solution proper motions for galaxies are determined as in the previous Section, 
using the same photometrically selected Sextans reference sample, the same spatial 
and proper-motion cuts, and with the number of surrounding reference stars set to 20.
In magnitude, we restrict the galaxy sample to within $I_C = 18.5 -22.5$.
The mean proper motion of galaxies (with respect to Sextans stars) is determined as 
a weighted mean, where the weights are given by the
proper-motion uncertainties of the galaxies, which dominate. 
However, instead of using the individual proper-motion uncertainties as determined from the scatter of the 
linear fits of positions with time, we use estimates adopted from the moving median of uncertainties with magnitude as shown 
in Fig.~\ref{fig:fig8}. 
This representation is statistically more sound, since many galaxies' uncertainties will have been underestimated due to chance 
alignment of the small number of average positions with time used in the relative proper-motion solution (see Section 6.2). 
The resulting value of the
weighted mean proper motion of 1276 galaxies with respect to Sextans stars is
$(\mu_{\alpha} , \mu_{\delta}) = (0.425\pm0.038, 0.061\pm0.038)$ mas~yr$^{-1}$. 
To these formal uncertainties in the weighted mean
we add in quadrature the uncertainty given by the scatter in the four Sextans star samples from Section 6.1.2. Thus, 
$(\mu_{\alpha} , \mu_{\delta}) = (0.425\pm0.051, 0.061\pm0.060)$ mas~yr$^{-1}$. 

\begin{figure}
\includegraphics[width=\columnwidth,angle=0.]{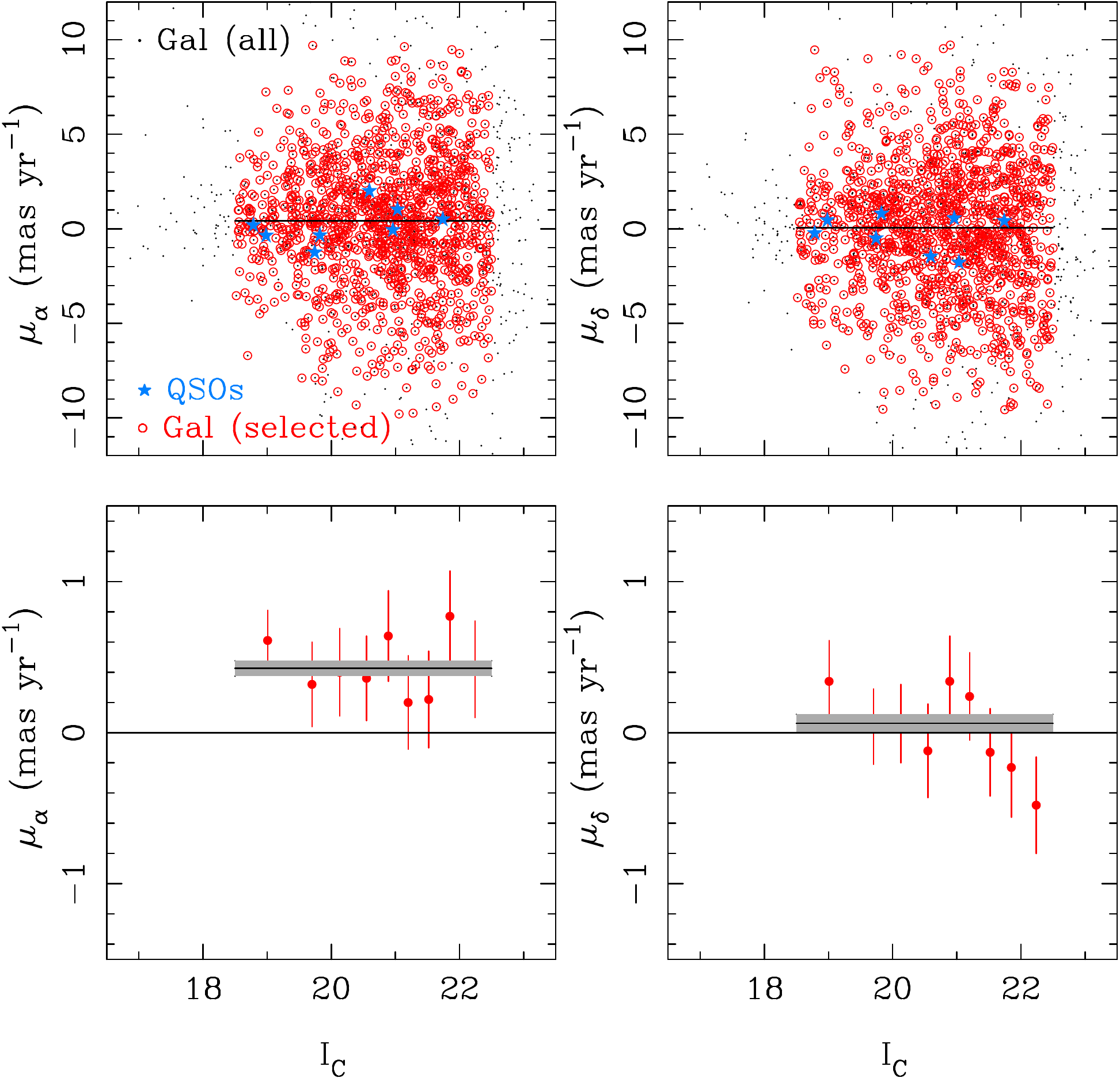}
\caption{
Local-solution proper motions of galaxies with respect to Sextans stars ($I_C=19-22$) as a function of 
magnitude. The top panels show individual proper motions of: all galaxies (dark points), galaxies selected for the 
purpose of mean determination (red circles), and QSOs (blue stars). The horizontal line shows the value of the 
weighted mean of galaxies. The bottom panels zoom in on the proper-motion scale, and show binned values for the selected 
sample of galaxies. The weighted mean is shown with a black line, and its formal uncertainty as a grey band.
}
\label{fig:fig12}
\end{figure}

In Figure~\ref{fig:fig12} we show the locally-corrected proper motions of external (distant) objects as a function of magnitude.
The top panels show individual proper motions of all galaxies, of galaxies selected for the mean determination 
(with the weighted mean indicated), and 
of eight quasars available in our study and extracted from the catalog of \citet{sou15}. 
The bottom panels show a magnified view of the mean-determining galaxies, but with the data binned. The 
value of the weighted mean proper motion of galaxies and its total uncertainty are represented with a dark line and
shaded band, respectively. No statistically significant trends with magnitude are seen. 
Inspection of similar plots of proper motions versus
color and versus $\xi$ and $\eta$ coordinates also do not show trends. 
The QSOs agree well with the galaxies, although their weighted mean has a formal uncertainty much larger than that of 
the galaxies: $(\mu_{\alpha} , \mu_{\delta}) = (0.064\pm0.235, -0.165\pm0.235)$ mas~yr$^{-1}$.

\section{Final Absolute Proper Motion of Sextans}

\begin{figure}
\includegraphics[width=\columnwidth,angle=-90.]{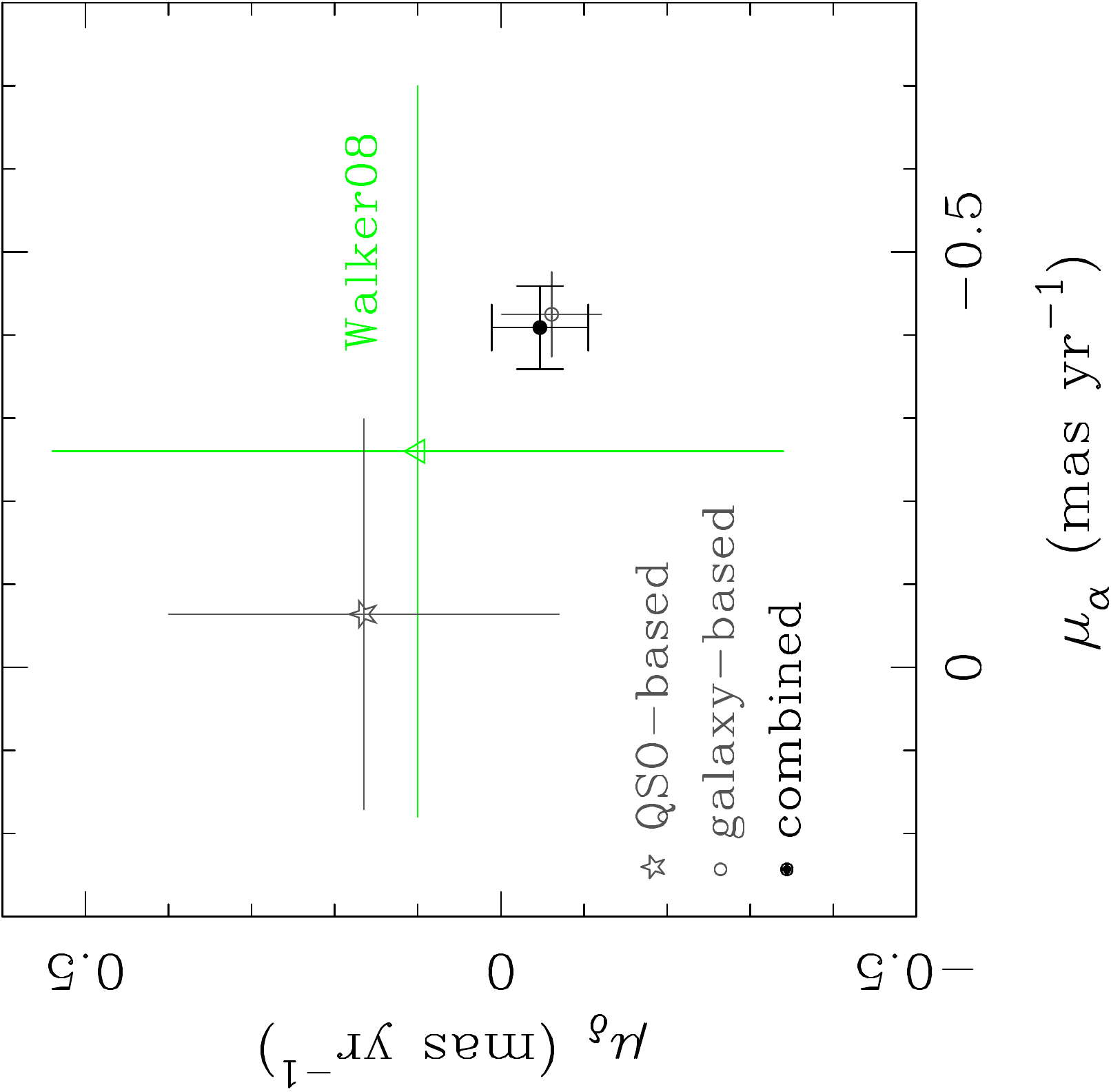}
\caption{
Absolute proper-motion determinations of Sextans. This study's three determinations are specified in the legend. The sole 
previous determination based on radial velocities alone \citep{wal08} is also labeled in the plot.
}
\label{fig:fig13}
\end{figure}

We adopt as the final proper motion of Sextans the weighted mean of the two determinations made in the previous Section
from galaxies and quasars.
The sign of the components must be flipped, so that the motion will reflect 
that of Sextans with respect to an inertial 
reference frame of extragalactic objects. The resulting value is: 
$(\mu_{\alpha} , \mu_{\delta}) = (-0.409\pm0.050, -0.047\pm0.058)$ mas~yr$^{-1}$. 
In Figure~\ref{fig:fig13} we show the measurements derived in this work, and the sole previous determination of
the proper motion of Sextans by \citet{wal08}. This latter measurement was not astrometric, but was derived from
radial velocities alone. While the estimate of \citet{wal08} has large uncertainties, it is bracketed by our two 
measurements based on QSOs and galaxies. 

\section{Velocity and Orbit}
Our derived proper motion is transformed to the Galactic rest frame by subtracting the solar reflex motion at the 
distance of Sextans.  For the Sun, we take the value of peculiar motion to be
$(U_{\odot},V_{\odot},W_{\odot}) = (-11.1,12.24,7.25)$ km~s$^{-1}$ 
\citep{sch10},
with positive $U$ radially away from the Galactic center, $V$ in the direction of Galactic rotation, 
and $W$ toward the North Galactic Pole. Uncertainties in these velocity components 
are of the order of $0.4 - 0.7$ km~s$^{-1}$, or a factor of ten times
smaller than those introduced by the measured proper motion; therefore, we ignore them here.
The adopted Local Standard of Rest (LSR) velocity is $238\pm8$ km~s$^{-1}$, and the Sun's distance
to the Galactic center assumed to be $R_{0} = 8.3\pm0.3$ kpc \citep{sch12}. While the uncertainty in the rotation of the 
LSR is comparable with that of the velocity of Sextans, we will not explore here its impact on 
the final velocity and orbit, since it is not crucial.

Sextans' spatial center is adopted from \citet{irw90}, heliocentric distance from \citet{oka17}, $d= 92.5\pm2.5$ kpc, 
and heliocentric radial velocity from \citet{wal09b}, $V_h = 224.3\pm 0.1$ km~s$^{-1}$.
We thus obtain Sextans' proper motion with respect to the Galactic rest frame: \\
$(\mu_{\alpha}^{GRF} , \mu_{\delta}^{GRF}) = (-0.303\pm0.050, 0.318\pm0.058)$ mas~yr$^{-1}$ in equatorial coordinates and, \\
$(\mu_l^{GRF} , \mu_b^{GRF}) = (-0.439\pm0.055, -0.028\pm0.053)$ mas~yr$^{-1}$ in Galactic coordinates.
The corresponding velocity components are: 
$(\Pi, \Theta, W) = (60\pm20, 196\pm22, 47\pm18)$ km~s$^{-1}$ in cylindrical coordinates at the location of Sextans, 
and $(U, V) = (197\pm23, 55\pm18)$ km~s$^{-1}$
along the Sun-to-Galactic-Center direction and orthogonal to it, respectively. 

Using these velocities an ensemble of orbits for Sextans is calculated in an analytic potential of the Milky Way, 
where the potential has three components and the halo potential is logarithmic \citep[e.g.,][]{dgv99}. 
Orbit parameter uncertainties are determined from Monte Carlo tests
where the initial conditions are varied based on the measurement uncertainties in the 
proper motion, heliocentric distance and radial velocity. Upper and lower values of these parameters represent
$1\sigma$ uncertainty ranges (specifically, the inner $68\%$ interval centered on the median of each distribution).  
In Table~\ref{tab:tab4} we present the derived orbital parameters, 
i.e., period, apo- and peri-Galactic distance, eccentricity, and inclination.
\begin{table}
\caption{Orbital parameters}
\label{tab:tab4}
\begin{tabular}{rrrrr}
\hline
\multicolumn{1}{c}{$P$} & \multicolumn{1}{c}{$r_a$} & \multicolumn{1}{c}{$r_p$} & \multicolumn{1}{c}{$ecc$} & \multicolumn{1}{c}{$\Psi$} \\
 \multicolumn{1}{c}{(Gyr)} & \multicolumn{2}{c}{(kpc)} & & \multicolumn{1}{c}{$(\degr)$} \\
\hline \\
$3.3^{+0.6}_{-0.4}$  & $132^{+26}_{-16}$  & $75^{+7}_{-12}$ & $0.28^{+0.04}_{-0.01}$ & $41^{+1}_{-1}$ \\
\hline
\end{tabular}
\end{table}

\begin{figure}
\includegraphics[width=\columnwidth]{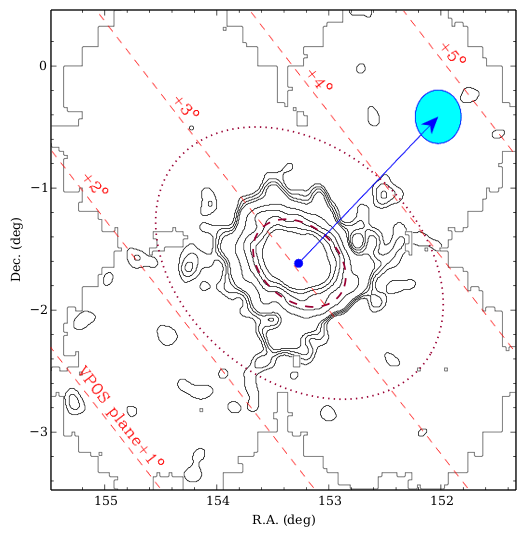}
\caption{Contour map of the Sextans stellar density reproduced with permission from \citet{rode16}, and determined 
from DECam photometry. The dashed and dotted ellipses
represent the core and tidal radii, respectively. The vector shows our measurement of the proper motion of Sextans 
(with respect to the Galactic rest frame), while the shaded ellipse represents the $1\sigma$ uncertainty. 
The diagonal, dashed lines show the orientation of and offset from the VPOS plane, as defined by \citet{paw15}.
Although Sextans' location in the sky is very close to the VPOS plane, its motion is away from it.
The jagged outer contour is indicative of the footprint of the DECam study. 
}
\label{fig:fig14}
\end{figure}

These results indicate that Sextans is currently moving toward apocenter, with its most recent pericenter 
passage some 0.4 Gyr ago. It has a low to moderate eccentricity, and a rather
modest inclination of the orbit plane with respect to the Galactic plane.
The coordinates of its orbit pole are $(l,b)=(239^{+10}_{-11}, -52^{+0.8}_{-0.3}) \degr$. 
This value is outside the range of the eight most concentrated
orbit poles as determined by \citet{paw13}, i.e., $(l, b) = (176\degr, -15\degr)$ 
with a radius of $\sim 30\degr$.  Their more recent determination \citep{paw15} based on the best-fit plane of the spatial 
distribution of 38 MW satellites, including newly discovered ones, is  $(l, b) = (164\fdg0, -6\fdg9)$.  
Based on our proper-motion measurement,
Sextans does not co-orbit with the vast polar structure (VPOS) 
characterized by \citet{paw13,paw15}, although spatially the dwarf galaxy does coincide with
this structure. Sextans' motion in the sky is roughly perpendicular to its major axis. \citet{rode16} 
examine a wide area around Sextans, using the Dark Energy Camera (DECam), and they find some overdensities
outside its nominal tidal radius, but no clear tidal tails. However, these overdensities seem to be aligned
along and perpendicular to its major axis, suggesting perhaps a morphological transition, as the galaxy
moves away from pericenter. In Figure~\ref{fig:fig14} we present the contour density map of Sextans 
from the study by \citet{rode16} (their Fig. 7) together with the proper-motion vector measured here.
The VPOS plane \citep[from][]{paw15}, with its pole at $(l,b) = (164\fdg0,-6\fdg9)$, is indicated
to illustrate just how close Sextans is to it.  Based on our measurement, 
Sextans is moving away from the VPOS plane, rather than parallel to it.

\section{Summary}
The absolute proper motion of the Sextans dwarf galaxy has been measured using three epochs of {\it Subaru} Suprime-Cam data spanning
$\sim 10$ years, thus providing the first astrometric measurement of this MW satellite. We carefully calibrate each epoch 
using {\it Subaru} dithered images and the {\it Gaia} DR1 catalog in the fields of Sextans and NGC 2419. Our proper-motion
study reaches a faint limit of $V = 24$ and covers an area of $26\farcm7\times23\farcm3$. The measured proper motion implies a
low eccentricity orbit with a period $\sim 3$ Gyr. The proper motion vector is perpendicular to the major axis of the satellite,
and seems to align with some of the stellar overdensities mapped by \citet{rode16} near its tidal radius. The proper-motion vector 
does not support Sextans' membership to the VPOS.

\section*{Acknowledgments}
This work was funded, in part, by the NASA Connecticut Space Grant College Consortium matched with equal funds from 
Southern Connecticut State University.
It makes use of data collected at the Subaru Telescope and obtained from SMOKA, which is operated by the 
Astronomy Data Center, National Astronomical Observatory of Japan.
This work has made use of data from the European Space Agency (ESA)
mission {\it Gaia} (\url{https://www.cosmos.esa.int/gaia}), processed by
the {\it Gaia} Data Processing and Analysis Consortium (DPAC,
\url{https://www.cosmos.esa.int/web/gaia/dpac/consortium}). Funding
for the DPAC has been provided by national institutions, in particular
the institutions participating in the {\it Gaia} Multilateral Agreement.
We also wish to thank Tammy Roderick for making available Figure 7 from their study.

%We thank the anonymous referee whose suggestions helped improve this manuscript.

%%%%%%%%%%%%%%%%%%%%%%%%%%%%%%%%%%%%%%%%%%%%%%%%%%



%\bibliographystyle{mnras}

%\begin{thebibliography}{99}
%\end{thebibliography}

%%%%%%%%%%%%%%%%%%%%%%%%%%%%%%%%%%%%%%%%%%%%%%%%%%

%%%%%%%%%%%%%%%%% APPENDICES %%%%%%%%%%%%%%%%%%%%%

%\appendix



%%%%%%%%%%%%%%%%%%%%%%%%%%%%%%%%%%%%%%%%%%%%%%%%%%

% Don't change these lines
\bsp	% typesetting comment
\label{lastpage}
\end{document}